\documentclass[journal]{vgtc}                     

\onlineid{0}

\vgtccategory{Research}

\title{Sharing Roughness with Hand-Outline Visualization\\ to Reduce Sensory Asymmetry in VR Collaboration}

\author{%
  \authororcid{Minju Baeck}{0000-0001-7179-2103},
  \authororcid{Yoonseok Shin}{0009-0004-4164-9559}, \authororcid{Hyunjin Lee}{0000-0002-4628-4921},  \authororcid{Boram Yoon}{0000-0003-3696-0145}, \authororcid{Sang Ho Yoon*}{0000-0002-3780-5350} and 
  \authororcid{Woontack Woo*}{0000-0002-5501-4421}
}

\authorfooter{
    \item
  	Minju Baeck, Yoonseok Shin is with KAIST UVR Lab.
  	E-mail: \{minjubaeck, sys7498\}@kaist.ac.kr
  \item
  	Hyunjin Lee, Boram Yoon is with KAIST PMRC.
  	E-mail: \{clairehj517, boram.yoon1206\}@kaist.ac.kr
  \item
  * Sang Ho Yoon is with KAIST HCI Tech Lab, and Woontack Woo is with KAIST UVR Lab and KAIST KI-ITC ARRC. 
  They are the corresponding authors.
  E-mail: \{sangho, wwoo\}@kaist.ac.kr
}

\abstract{%
In collaborative VR, asymmetric access to haptic hardware creates a critical information gap: tactile evidence remains private to the haptic user, hindering the shared understanding needed for joint decision-making. While prior work has explored crossmodal sensory cues in virtual environments, it remains unclear how such cues should be designed for asymmetric collaboration, where collaborators receive information through different modalities. In our setting, the haptic user feels roughness through fingertip vibration, whereas the non-haptic user relies on vision alone. To reduce this asymmetry, we propose externalizing an object’s tactile state through a glanceable hand-outline visual proxy. Specifically, we examine whether abstract visual roughness cues based on line shape and motion can encode three discrete roughness levels for both haptic and non-haptic users. Two preliminary studies establish a shared visual semantics by identifying visually distinguishable cues for non-haptic users and validating their visuo-haptic correspondence for haptic users. In a main study of a collaborative sorting task, showing this visualization on both users’ hands significantly reduced completion time relative to a no-visualization baseline. \replaced{Moreover, NU-side cue visibility was associated with higher confidence
and perceived contribution for the non-haptic user.}{Moreover, showing the cue on the non-haptic user’s own hand, rather than only on the partner’s hand, significantly increased perceived contribution and confidence.} These findings show that hand-anchored abstract visual cues provide a lightweight means of externalizing object tactile state, reducing information asymmetry without compromising social presence. 
}

\keywords{Asymmetric Collaboration, Collaborative Virtual Reality, Sensory Asymmetry, Roughness Perception, Tactile Visualization}

\teaser{
  \centering
  \includegraphics[width=\linewidth, alt={A view of clouds with orange sunrays shining through from behind.}]{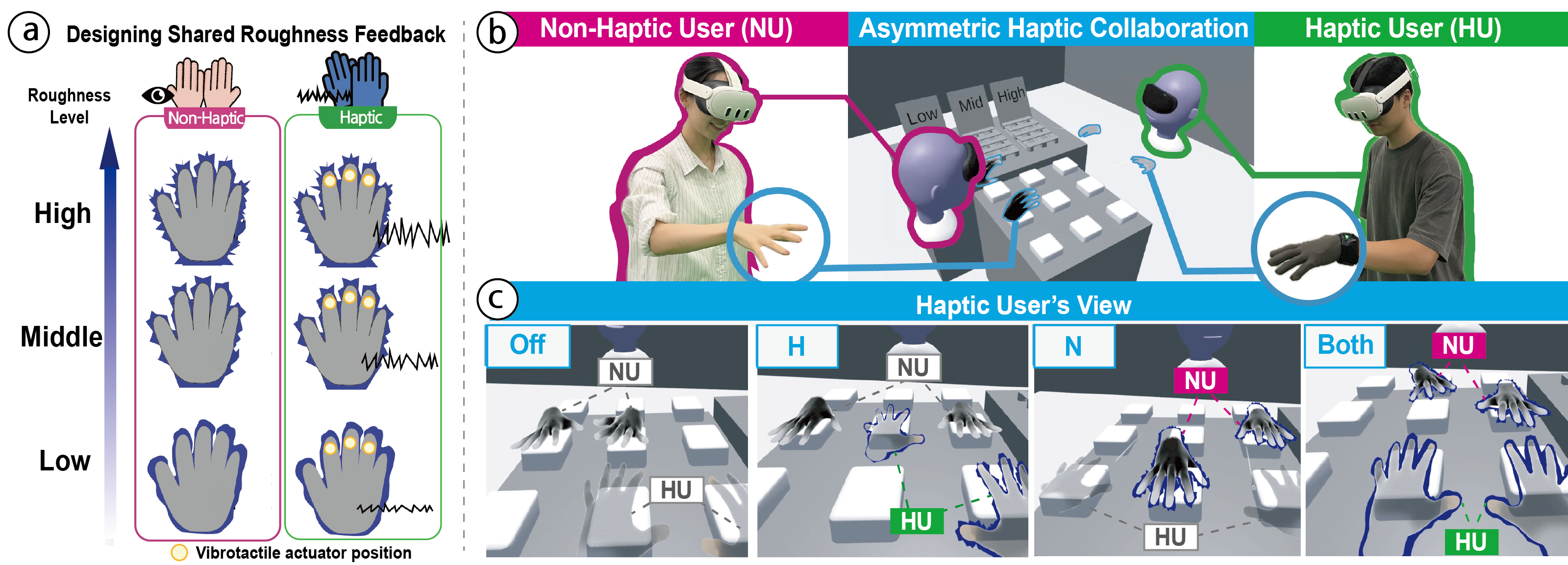}
  \caption{Study overview.
  (a) Preliminary studies validating line shape and motion as visual cues for roughness.
  (b) Asymmetric haptic collaboration setup with a haptic user (\textbf{HU}) and a non-haptic user (\textbf{NU}).
  (c) Main study visibility modes: \textbf{Off} (no visualization), \textbf{H} (on the HU’s hand), \textbf{N} (on the NU’s hand), and \textbf{Both} (on both hands).}
  \label{fig:teaser}
}

\graphicspath{{figs/}{figures/}{pictures/}{images/}{./}} 

\usepackage{times}                     
\usepackage{verbatim}                  
\usepackage{multirow}

\usepackage{booktabs}                  
\usepackage{lipsum}                    
\usepackage{mwe}                       
\usepackage{ccicons}                   

\usepackage{mathptmx}                  
\usepackage{verbatim}                  

\usepackage[table]{xcolor} 
\definecolor{darkgreen}{RGB}{69, 169, 79} 
\definecolor{PRStrong}{RGB}{255,224,224} 
\definecolor{PRMiddle}{RGB}{224,234,255} 
\definecolor{PRWeak}{RGB}{224,245,224}   
\newcommand{\prS}[1]{\cellcolor{PRStrong}#1}
\newcommand{\prM}[1]{\cellcolor{PRMiddle}#1}
\newcommand{\prW}[1]{\cellcolor{PRWeak}#1}
\usepackage{tabu}                      
\usepackage{amsmath}
\usepackage{graphicx}

\usepackage{tabularx,pifont}

\newcolumntype{L}[1]{>{\raggedright\arraybackslash}p{#1}}
\usepackage{tabularx}
\usepackage{array}
\newcolumntype{Y}{>{\raggedright\arraybackslash}X}

\usepackage{enumitem} 

\usepackage[final,commandnameprefix=ifneeded]{changes}
\setdeletedmarkup{}

\begin{document}


\firstsection{Introduction}
\maketitle

Virtual Reality (VR) enables users to collaborate through direct interaction with virtual objects. Recent advances in consumer haptic gloves and hand tracking enrich this process by conveying properties such as roughness, surface condition, and other interaction-related tactile sensations. In collaborative VR, haptic feedback has been shown to enhance immersion, teamwork, and task performance~\cite{sasaki2025exploring,tian2023group,venkatraj2024shareyourreality}. This capability is especially valuable in collaborative scenarios such as \replaced{design review, material or finish comparison, interactive product experiences, and haptic playtesting, where collaborators inspect virtual objects and need a shared reference for coarse tactile qualities.}{design review, industrial evaluation, and interactive product experiences, where collaborators need to compare and discuss an object’s texture, surface quality, or material condition.} In such scenarios, the goal is often not detailed tactile replication, but rapid discrimination \replaced{and discussion of coarse tactile qualities.}{of tactile qualities.} In practice, however, not all collaborators have access to haptic hardware. As a result, task-relevant tactile information may only be available to limited users, creating an asymmetric collaboration setting. We use asymmetric collaboration to refer to collaborative VR settings in which users have unequal access to devices, roles, or task-relevant information~\cite{chi2025l,tong2023towards}.

This asymmetry is especially problematic when collaborators jointly assess tactile qualities such as roughness or surface condition. While the haptic user can directly access these cues, the non-haptic user must infer them through observation or conversation. When tactile information is only available to one partner in asymmetric VR collaboration, an information gap can emerge, potentially leading to unequal participation, reduced shared control, and miscommunication~\cite{chi2025l}. One possible way to address this gap is to externalize object-related tactile information through visual cues. In collaborative settings, visual cues have been widely studied for reducing ambiguity and supporting communication~\cite{bovo2022cone,jing2022impact, kim2023visualizing}.

Tactile perception is shaped by multisensory integration, and tactile qualities may therefore be conveyed not only through haptic hardware but also through perception-driven visual cues that bias or support sensory interpretation~\cite{di2022roughness,ricci2024perception}. In Extended Reality (XR) interaction, the virtual hand can serve as a natural channel for such cues while also enhancing immersion and presence~\cite{sanchez2005presence}. Prior work has used hand-related visual cues in VR, including color, abstract form, and motion, to represent affective and interaction-related sensory information such as thermal, pain, and force cues~\cite{kim2023visualizing,kocur2023effects,martini2013color}, as well as material qualities such as roughness and slipperiness~\cite{baeck2025visuo,suzuishi2020visual}. Among tactile object properties~\cite{tymms2018quantitative}, roughness is a fundamental and perceptually salient attribute, making it a suitable target for studying how tactile information can be externalized and shared. Extending this work from individual sensory modulation to asymmetric collaboration, we investigate whether hand-centered roughness cues can \replaced{externalize task-relevant tactile state for shared use between collaborators.}{communicate tactile information between collaborators.}

In collaborative settings, the issue is not only how information is encoded, but also who has access to it. Prior work in collaborative VR suggests that access to visual information can influence coordination, performance, and participation balance~\cite{jing2022impact,khokhar2025enhancing,sasaki2025exploring, yoon2020evaluating}. 
This raises the possibility that cue visibility is itself an important collaboration design variable. In particular, whether the cue is shown to the haptic user, the non-haptic user, both users, or neither user may shape access to task-relevant tactile information and participation during collaboration.

In this paper, we use roughness as a representative tactile property \deleted{of virtual objects }during hand--object interaction to study how object-related tactile information can be externalized for shared use in asymmetric collaboration. Rather than aiming for high-fidelity tactile reproduction, we support rapid comparison and shared understanding through a glanceable, contact-situated cue that conveys coarse categories such as smooth, medium, and rough. To this end, we externalize coarse \deleted{object }roughness information through hand-anchored visual proxies whose shape and motion convey three \deleted{discrete }roughness levels. 

To realize this approach, we address two questions: whether a hand-anchored cue is visually distinguishable to non-haptic users while remaining \replaced{aligned}{consistent} with the haptic user’s tactile \replaced{state}{experience}, and how its visibility affects asymmetric collaboration. We therefore conduct two preliminary studies to establish a shared \replaced{visual--haptic roughness mapping for the main study}{visuo-haptic roughness code}, followed by a main study comparing four cue-visibility configurations: \textit{Off} (the cue is shown to neither user), \textit{H} (the cue is shown only on the haptic user’s hand), \textit{N} (the cue is shown only on the non-haptic user’s hand), and \textit{Both} (the same cue is shown on both users’ hands). Across all conditions, the cue externalizes the haptic user’s current tactile state.

The contributions of this work are threefold:
\begin{enumerate}[leftmargin=*, itemsep=2pt, topsep=2pt]
  \item \textbf{A lightweight hand-anchored visualization method} for asymmetric haptic collaboration that externalizes object roughness through dynamic hand-outline cues, restoring shared access to task-relevant tactile information without additional hardware.
  
    \item \textbf{A validated visual--haptic roughness mapping} for \replaced{sharing task-relevant coarse roughness}{collaborative roughness judgment}, established through two studies of visual discriminability and visuo-haptic correspondence.
    
    \item \textbf{An empirical account of cue placement in asymmetric collaboration}, showing how non-haptic-user cue visibility shapes efficiency, participation, and workload equity, and informing tactile-state visualization design in collaborative VR.
\end{enumerate}

\section{Related Work}

\subsection{Asymmetric Collaboration and Haptic Experience}

Asymmetry is a recurring challenge in collaborative VR systems. Prior work has shown that differences in devices, viewpoints, representations, and interaction capabilities shape information access and participation across collaborators~\cite{chi2025l,kim2023visualizing,kim2019evaluating,piumsomboon2019effects,shin2022effects,yoon2023effects}. \added{Collaborative VR and virtual-world systems have been used for design review and spatial design communication, enabling stakeholders to inspect, discuss, and revise design alternatives in shared 3D environments~\cite{koutsabasis2012value,prabhakaran2022bim}. However, these systems primarily support visual and spatial communication; how tactile material states such as roughness can be shared under uneven haptic access remains less explored.} Across a range of collaborative contexts, prior work has also explored multi-user haptic systems that share object properties such as roughness or weight, as well as learning and training systems that support shared haptic experience~\cite{van2023haptic,webb2022haptic}. Such haptic sharing has been shown to benefit task performance and social presence~\cite{nam2008roles,sallnas2000supporting,sasaki2025exploring,tong2023towards}, demonstrating the collaborative value of sharing object-related haptic information. However, these systems typically assume that multiple users have direct haptic access. In contrast, we focus on asymmetric collaboration in which only one user has haptic access. Prior work has rarely examined how object properties can be shared under unequal haptic access, or how non-haptic collaborators can access that information.

\subsection{Roughness as a Crossmodal Attribute}

Roughness is a fundamental tactile attribute that, although primarily conveyed through touch, can also be shaped by visual cues~\cite{di2022roughness,heller1982visual,okamoto2012psychophysical}. In haptics research, roughness is often rendered through vibrotactile stimulation, as the fingertip is highly sensitive to vibration and variations in stimulation parameters can produce distinct roughness levels~\cite{asano2014vibrotactile,baeck2025visuo,gonzalez2014analysis,hollins2001vibrotactile}. Prior work further suggests that roughness can be evoked through crossmodal cues rather than only through physical simulation~\cite{ricci2024perception}. For example, perceived roughness can be modulated by visual motion and contour shape~\cite{baeck2025visuo,blazhenkova2018angular,suzuishi2020visual,ujitoko2019modulating}.

While some VR haptics research aims for high-fidelity tactile reproduction, practical interaction may also be supported by a small number of perceptually distinguishable levels. Prior work suggests that haptic perception can exhibit categorical structure, such that continuous tactile variation may be perceived as a small number of discrete perceptual groups~\cite{gaissert2012haptic}. Related studies further show that visuo-haptic integration can support meaningful roughness distinctions even with only two underlying haptic levels, and that pseudo-haptic feedback can convey tactile-like texture properties through distinguishable levels without haptic devices~\cite{gunther2022smooth,sato2020modifying}. Together, these findings motivate our focus on coarse, perceptually separable roughness categories rather than detailed tactile replication. However, it remains unclear whether crossmodal cues can support asymmetric collaboration, rather than only individual perception, across users with different sensory access.

\subsection{Externalizing Haptic States through Shared Visual Cues}

Collaborative VR research suggests that shared visual cues can support coordination by exposing otherwise implicit partner states. Prior work has explored sharing eye gaze, hand gestures, pointers, and sketches to enrich nonverbal communication and task coordination~\cite{bovo2022cone,jing2022impact,khokhar2025enhancing}. Kim et al. visualized a worker’s hand force for a remote expert and showed that making this hidden state visible improved awareness of collaborator force and object weight, while also increasing social presence~\cite{kim2023visualizing}. Related work on collaborative manipulation similarly suggests that effective coordination depends not only on access to one’s own actions, but also on access to collaborators’ interaction states~\cite{kim2024providing,wang2025effects}.

However, two issues remain underexplored in asymmetric haptic collaboration: whether tactile-state cues are interpretable to non-haptic users, and how access to those cues should be distributed across collaborators. These questions matter because visualization design in collaborative VR can shape both social presence and collaboration outcomes: avatar visibility affects social presence, with its effects on coordination and performance depending on task context~\cite{wang2025effects,yoon2023effects}, while task-specific cues such as hand-force visualization can improve awareness of task-relevant interaction states~\cite{kim2023visualizing}. In VR, the virtual hand provides a natural channel for such cues while also supporting immersion and presence~\cite{sanchez2005presence}. Prior work has used hand-related visual cues, including color, form, and motion, to represent sensory information such as thermal, pain, force, roughness, and slipperiness~\cite{baeck2025visuo,kim2023visualizing,kocur2023effects,martini2013color,suzuishi2020visual}. Compared with detached textual or numeric readouts, hand-anchored cues can present tactile information in a contact-situated and perceivable form, making them suited for in-situ distribution during collaboration.

\section{Translating Haptic\deleted{s}  Cues into \deleted{Shared }Visual Cues}

\begin{figure}[t]
  \centering
  \includegraphics[width=0.9\columnwidth]{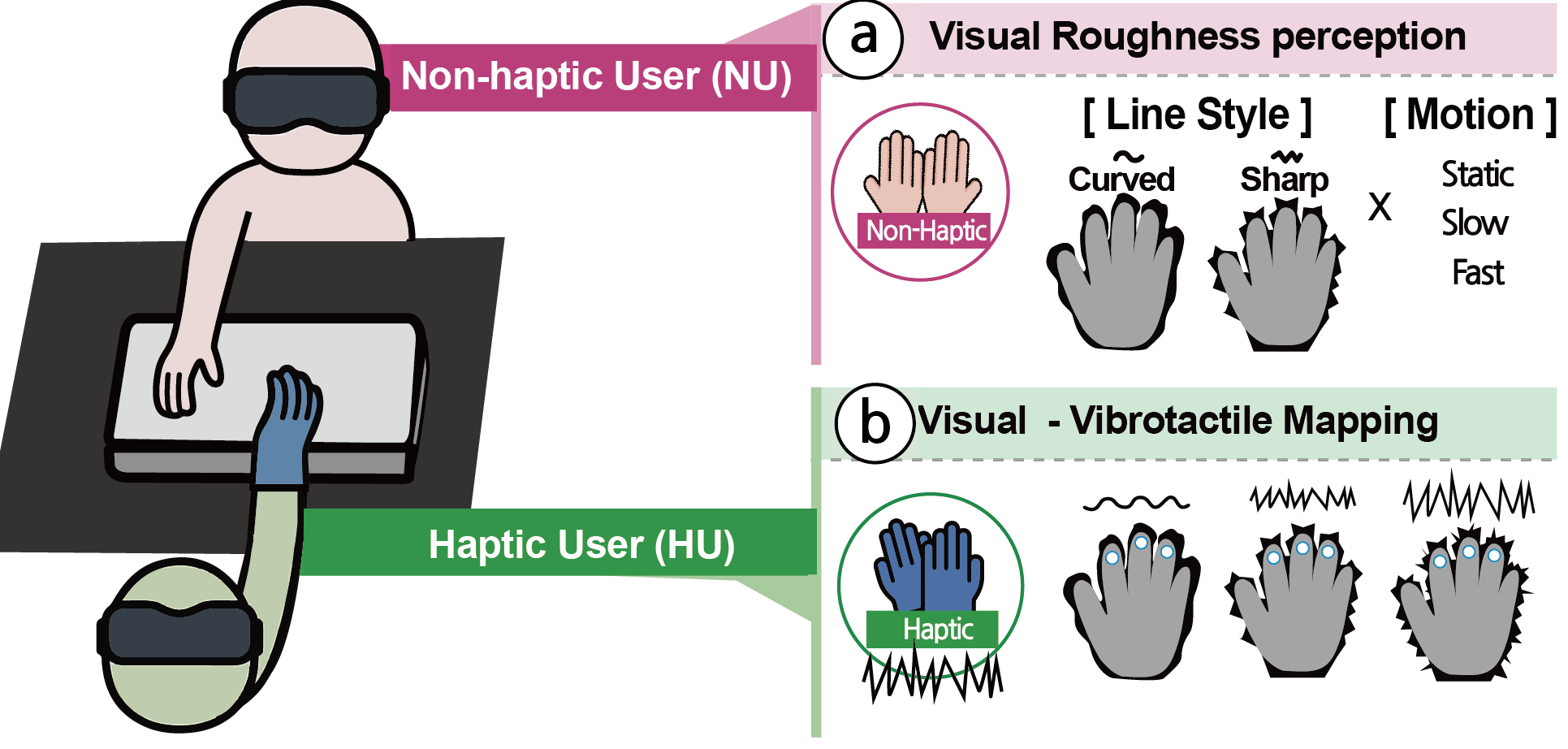}
  \caption{Study setup for preliminary studies. (a)~Non-haptic users discriminate visual roughness cues. (b)~Haptic users verifying their alignment with vibrotactile feedback.}
  \label{fig:preover}

\end{figure}

\added{The preliminary studies establish the visual--haptic roughness code used in the main collaborative study. For brevity, we use HU to denote the haptic user and NU to denote the non-haptic user.}
To mitigate information asymmetry in haptic hardware-asymmetric VR collaboration, we \replaced{externalize the HU's}{introduce a shared visual channel that} 
 tactile information through hand-anchored cues. Before \deleted{evaluating its collaborative effects in }the main study (see~\cref{subsec:4}), we first \deleted{establish and} validate this channel through two \deleted{sequential} research questions (see~\cref{fig:preover}).


\subsection{Design Space of Roughness}

Our scenario targets rapid roughness discrimination rather than detailed tactile reproduction. For the haptic user, we used fingertip vibrotactile feedback, as the fingertip is highly sensitive to vibration and vibrotactile parameters can convey distinct roughness levels~\cite{asano2014vibrotactile,baeck2025visuo,gonzalez2014analysis,hollins2001vibrotactile}. To communicate roughness to non-haptic users from a crossmodal perspective, we adopted a perception-driven design approach, which suggests that tactile qualities in virtual interfaces can be evoked through perceptual cues~\cite{ricci2024perception}. Based on this perspective, we designed the visualization around two cue dimensions for conveying coarse roughness during hand--object contact: \textit{Line Shape} and \textit{Motion}. \deleted{For brevity, we use NU to denote the non-haptic user and HU to denote the haptic user.} We then ask whether these cues are visually distinguishable to the NU and whether they can be meaningfully mapped to perceived roughness by the HU.

\noindent\textbf{RQ1-1.} Can the non-haptic user (NU) reliably discriminate the proposed cue variants?

\noindent\textbf{RQ1-2.} Can the haptic user (HU) consistently map these cues to perceived tactile roughness?

\subsubsection{Hand-Outline Visualization for Roughness}
\label{subsec:vib}

To visually convey roughness to non-haptic users during hand--object interaction in VR, we adopted an animated hand outline as a visual proxy for roughness information. Prior work by Baeck et al. used \textit{Line Shape} and \textit{Motion} as abstract visual parameters in visuo-tactile feedback to study affective responses to roughness-related stimulation~\cite{baeck2025visuo}. At the same time, earlier crossmodal findings suggest that both line shape and visual motion can influence roughness impressions, making them plausible candidates for visual roughness representation~\cite{blazhenkova2018angular,suzuishi2020visual,ujitoko2019modulating}. Because Baeck et al. examined these parameters in combination with vibrotactile feedback, we adopted the same parameterization of \textit{Line Shape} and \textit{Motion} to test whether it could also support roughness discrimination through visual cues alone. We considered both dimensions because their combination offered a plausible way to represent multiple roughness levels rather than a single binary contrast. Specifically, we combined \textit{Line Shape} (\textit{Curve}, \textit{Sharp}) with \textit{Motion} (\textit{Static}, \textit{Slow}, \textit{Fast}), yielding six visualization conditions plus a \textit{None} baseline (see~\cref{tab:visual_conditions}). We chose the hand as the visualization channel because it naturally anchors hand--object interaction and remains visible to both users. To reduce distraction, we used a translucent gray hand representation~\cite{voisard2023effects}, consistent with prior work suggesting that high visual fidelity is not required for agency or body ownership~\cite{argelaguet2016role,kilteni2012sense}. The resulting outline remains contact-situated, glanceable, and minimally intrusive.

\textbf{Implementation.} We implemented these styles using a modified version of the Quick Outline Unity asset\footnote{\url{https://assetstore.unity.com/packages/tools/particles-effects/quick-outline-115488}}, applying separate shaders to the outline and the fill. The outline shader introduced time-dependent variations in vertex positions, parameterized by \replaced{a shader wave-speed parameter ($s_w$) and a shape function. Following Baeck et al.~\cite{baeck2025visuo}, \cref{tab:visual_conditions} reports the $s_w$ values for the Slow/Fast settings; Static applied no time-dependent deformation. Because Curve and Sharp used different deformation functions, $s_w$ values are style-specific and not directly comparable across line shapes.}{motion speed and a shape function, producing the intended visual effects of line shape and motion. The motion-speed values followed the implementation of Baeck et al.~\cite{baeck2025visuo}.}

\begin{table}[t]
    \centering
    \caption{Visual conditions by \textit{Line Shape} and \textit{Motion}. Parentheses denote shader wave-speed $s_w$; Static applied no time-dependent deformation.}
    \label{tab:visual_conditions}
    \resizebox{0.8\linewidth}{!}{%
    \begin{tabular}{lll}
    \toprule
    \textbf{Line Shape} & \textbf{Motion ($s_w$)} & \textbf{Color} \\
    \midrule
    None   & --                   & -- \\
    Curve  & Static / Slow (2.0) / Fast (10.0) & Gray \\
    Sharp  & Static / Slow (0.25) / Fast (2.0) & Gray \\
    \bottomrule
    \end{tabular}
    }
    
\end{table}

\subsubsection{Tactile Feedback Design for Roughness} 
\label{sec:tactilefeedback}

Our collaboration scenario focuses on an asymmetric setting in which one user wears a vibrotactile glove while the other has no haptic access. For tactile feedback, we adopted the same vibrotactile parameterization as Baeck et al.~\cite{baeck2025visuo}, using the bHaptics TactGlove DK2~\cite{bHapticsTactGlove} to deliver stimulation. Stimuli were applied simultaneously to the index, middle, and ring fingers to approximate tactile sensations during hand--object contact. Following prior findings that roughness perception increases with vibrotactile intensity and on-duration~\cite{hollins2000imposed}, Baeck et al. combined two \replaced{bHaptics intensity settings (10 and 80)}{intensity levels (10\% and 80\%)} with four on-durations ($5$, $20$, $30$, and $40\,\mathrm{ms}$), yielding eight conditions. Their preliminary study confirmed that these conditions were \deleted{reliably }distinguishable. Based on these validated results, we adopted three representative levels (Low: I10\_D30, Middle: I80\_D5, High: I80\_D30) as tactile conditions without \deleted{conducting }an additional pilot study. \added{The TactGlove DK2 uses LRA actuators with a 170\,Hz resonant frequency and 1G peak acceleration at 100\% intensity.} \replaced{I10/I80 denote bHaptics Designer intensity settings of 10 and 80 on a 0--100 scale, not vibration frequencies, and D5/D30 denote vibration on-durations of 5\,ms and 30\,ms.}{Here, I10/I80 indicate 10\% and 80\% intensity, respectively, and D5/D30 denote vibration on-durations of 5\,ms and 30\,ms.}

\subsection{
Preliminary Study~1: Visual Discriminability for Non-haptic Users}
\label{subsec:ps1}
\added{Preliminary Study~1 tested whether the candidate visual cues were sufficiently distinguishable to serve as shared roughness labels for the main study.}
To address \textbf{RQ1-1}, we first test\added{ed} whether non-haptic users can reliably discriminate the proposed visual cue variants with brief, glanceable viewing, which is a prerequisite for using the cue as a shared tactile-state signal in collaboration.

\begin{figure}[t]
  \centering
  \includegraphics[width=\columnwidth]{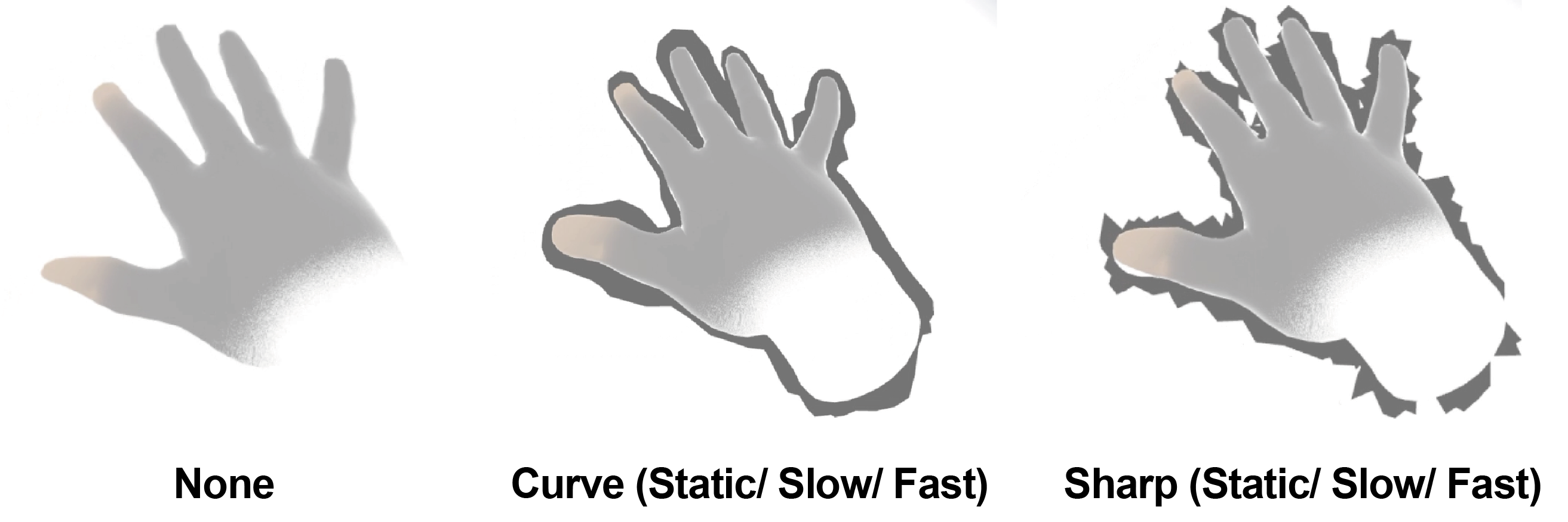}
  \caption{Visual roughness feedback stimuli: \textit{None}, and \textit{Line shape} (\textit{curve}, \textit{sharp}) combined with \textit{Motion} (\textit{static}, \textit{slow}, \textit{fast}).}
  \label{fig:visustimuls1}
\vspace{-0.3cm}
\end{figure}

\textbf{Stimulus.} Seven visual conditions were created by combining \textit{Line Shape} (\textit{curve}, \textit{sharp}) and \textit{Motion} (\textit{static}, \textit{slow}, \textit{fast}), along with a None baseline (see~\cref{fig:visustimuls1}). The VR environment was built in Unity 6000.1.f1 and run on a Meta Quest 3 headset. \added{In the \textit{slow} and \textit{fast} conditions, the outline animated continuously at different speeds, whereas in the \textit{static} condition, it remained fixed, presenting only the \textit{curve} or \textit{sharp} shape without motion.}

\textbf{Participants and Procedure.} Sixteen participants (7 male, 9 female; age 23–34 years, $M$ = 27.6) took part in the study. Most had extensive VR experience, with 14 having used VR devices more than 10 times; only two reported fewer than five experiences. When participants touched a block, an outline appeared around the hand.

\begin{figure}[t]
  \centering
  \includegraphics[width=\columnwidth]{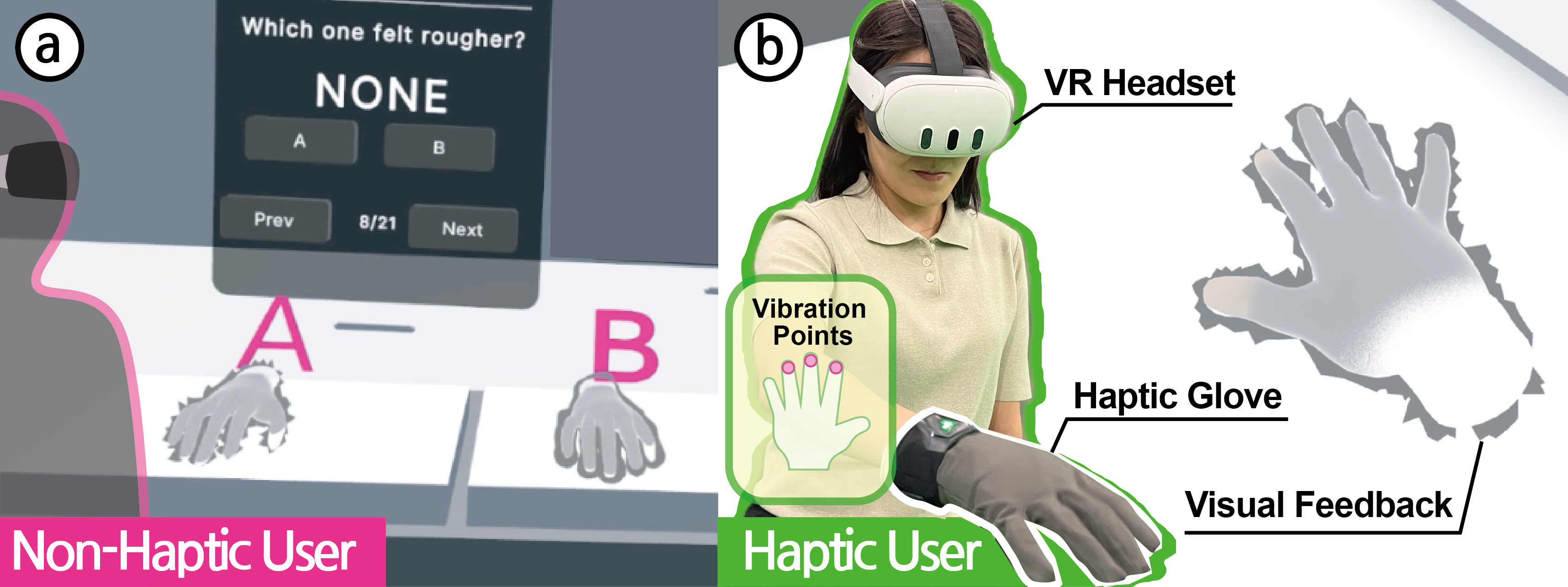}
\caption{System setup. a) Non-haptic user judging roughness from hand–outline visualizations (A/B test). b) Haptic user with VR headset and TactGlove DK2; contact with a virtual object triggers simultaneous fingertip vibrotactile and visual feedback.}
\label{fig:systemsetup}
\vspace{-0.3cm}
\end{figure}

Participants completed four blocks in a single session. Each block consisted of 21 trials\deleted{,} covering all unordered pairs of the seven conditions ($\binom{7}{2}=21$), \replaced{with randomized trial order and left/right A--B placement.}{with trial order randomized within each block and left/right A--B placement randomized across trials.} As shown in \cref{fig:systemsetup}-(a), two stimuli were presented simultaneously in each trial. \deleted{In the \textit{slow} and \textit{fast} conditions, the outline animated continuously, whereas in the \textit{static} condition, it remained fixed to present the \textit{curve} or \textit{sharp} shape without motion.} Participants indicated which stimulus appeared rougher (A or B). In the final block, they additionally rated the perceived roughness of both stimuli on a 1--100 scale (\replaced{higher = rougher}{higher values indicating greater roughness}) and reported their confidence \added{in each judgment} on a 5-point Likert scale (1 = not confident\deleted{at all}, 5 = very confident). The study was approved by the IRB, and all participants provided informed consent. Participants received approximately \$8~USD.

\begin{figure}[t]
  \centering
  \includegraphics[width=0.9\linewidth]{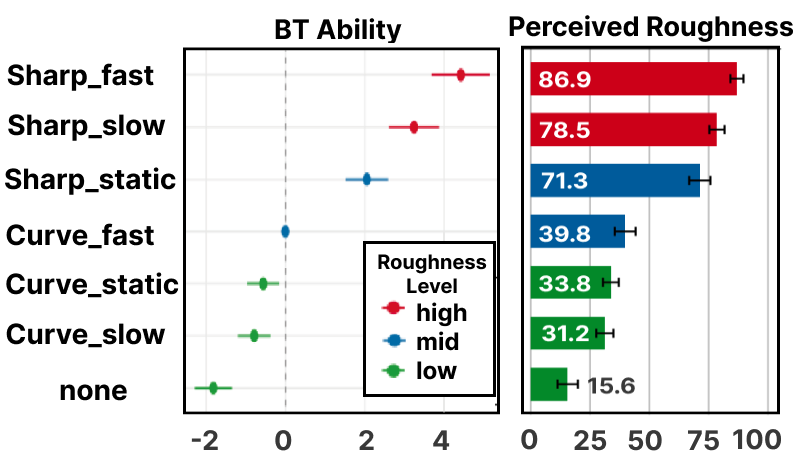}
  \vspace{-0.3cm}
  \caption{Preliminary Study 1 results. Left: BT ability estimates with 95\% CIs. Right: final-block mean roughness ratings (1--100; higher = rougher) with SE error bars. Colors indicate BT-based roughness tiers; \emph{none} is treated as a baseline.}
  \label{fig:bt_ci_main}
  \vspace{-0.3cm}
\end{figure}

\textbf{Analysis.} \replaced{We analyzed seven visual stimuli: six \textit{Line Shape} $\times$ \textit{Motion} combinations and the \emph{none} baseline.}{We analyzed seven visual stimuli (T1–T3).} For all 21 unordered pairs ($\binom{7}{2}$), A/B sides were collapsed and exact binomial tests against chance ($H_0:p=0.5$) were performed with Holm correction; Wilson 95\% CIs are reported. Practical equivalence was assessed \replaced{with}{using} two one-sided tests (TOST) with a ROPE of $\pm0.15$, classifying each pair as \emph{Discriminable} (Holm $p<.05$), \emph{Equivalent} (both TOST $p<.05$), or \emph{Inconclusive}. \replaced{We then fit a Bradley--Terry (BT) model to the collapsed win--loss counts, applying a 0.5 continuity correction for near-unanimous pairs. We report ability parameters $\beta$ (log-odds, larger = rougher) with Wald 95\% CIs and normalized strengths $s=\exp(\beta)/\sum\exp(\beta)$. \cref{fig:bt_ci_main} visualizes the BT estimates sorted by roughness, with a dashed reference at $\beta=0$ and color-coding for High/Middle/Low; \emph{none} is treated as a baseline.}{We then fit a Bradley--Terry (BT) model to the win--loss counts (A/B collapsed), applying a 0.5 continuity correction for near-unanimous pairs. We report ability parameters $\beta$ (log-odds, larger = rougher) with Wald 95\% CIs, and a normalized strength $s=\exp(\beta)/\sum\exp(\beta)$ for readability. \cref{fig:bt_ci_main} visualizes the BT estimates with confidence intervals, sorted by roughness, with a dashed reference at $\beta=0$ and color-coding for High/Middle/Low; \emph{none} is treated as a baseline only.} \added{Final-block roughness ratings were analyzed with a Friedman test across stimuli, and confidence was summarized descriptively at the pairwise-judgment level.}

\begin{table}[t]
\centering
\caption{Non-discriminable pairs in Preliminary Study~1. Win rates include Wilson 95\% CIs ($n=48$).}
\small
\begin{tabular}{lcc}
\toprule
\textbf{Pair} & \textbf{Win rate [95\% CI]} & \textbf{Decision} \\
\midrule
\textit{curve\_slow} vs.\ \textit{curve\_static} & 0.50 [0.36, 0.64] & \textbf{Equivalent} \\
\textit{curve\_fast} vs.\ \textit{curve\_static} & 0.60 [0.46, 0.73] & \textbf{Inconclusive} \\
\bottomrule
\end{tabular}
\label{tab:ab_summary_main2}
\vspace{-0.3cm}
\end{table}

\textbf{Results.} Across the 21 pairs, 19 differed from chance after Holm correction, indicating robust visual discriminability. \replaced{As summarized in \cref{tab:ab_summary_main2}, \emph{Sharp} variants were consistently judged rougher than \emph{Curve} variants and the \emph{None} baseline. Within-family differences were clearer for \emph{Sharp} than for \emph{Curve}: the only two non-discriminable pairs both occurred within the \emph{Curve} family.}{All contrasts involving a \emph{Sharp} stimulus against \emph{Curve} or \emph{None} favored \emph{Sharp}. Within-family orderings also emerged: for \emph{Sharp}, fast $>$ slow $>$ static (all Holm-adjusted $p \le .001$); for \emph{Curve}, only \emph{curve\_fast} $>$ \emph{curve\_slow} was reliable ($p=1.07\times10^{-4}$). TOST equivalence tests indicated that \emph{curve\_slow} vs.\ \emph{curve\_static} was statistically equivalent (win rate = 0.50, 95\% CI [0.36, 0.64]; $p_{\max}=0.023$), whereas \emph{curve\_fast} vs.\ \emph{curve\_static} remained inconclusive (0.60, 95\% CI [0.46, 0.73]; Holm $p=.387$; $p_{\max}=0.300$).}

The BT model placed all stimuli on a common latent \added{roughness} scale: \emph{sharp\_fast} $>$ \emph{sharp\_slow} $>$ \emph{sharp\_static} $>$ \emph{curve\_fast} $>$ \emph{curve\_static} $>$ \emph{curve\_slow} $>$ \emph{none} (see ~\cref{fig:bt_ci_main}). \added{Final-block roughness ratings further supported the BT-based ordering. A Friedman test showed a significant effect of stimulus, $\chi^2(6)=81.08$, $p<.001$, Kendall's $W=.845$, with mean ratings following the same order as the BT estimates (see~\cref{fig:bt_ci_main}). Descriptively, confidence was high overall ($M=4.11/5$), but lower for within-\emph{Curve} comparisons ($M=2.90/5$) than for other pairs ($M=4.32/5$), consistent with weaker separability among curved styles.}


\replaced{Based on these results, we grouped the visual stimuli into three roughness tiers while treating \emph{none} as a baseline: High=\{\emph{sharp\_fast}, \emph{sharp\_slow}\}, Middle=\{\emph{sharp\_static}, \emph{curve\_fast}\}, and Low=\{\emph{curve\_static}, \emph{curve\_slow}\}. To account for weaker separability among the \emph{Curve}-family stimuli near the Middle--Low boundary, Preliminary Study~2 used overlapping candidate pools for haptic--visual matching (see~\cref{tab:vh_mapping}).}{For \cref{subsec:ps1}, we discretized this continuum into three tiers: \textbf{High}=\{ \emph{sharp\_fast}, \emph{sharp\_slow}\}, \textbf{Middle}=\{\emph{sharp\_static}, \emph{curve\_fast}\}, \textbf{Low}=\{\emph{curve\_static}, \emph{curve\_slow}\}, with \emph{none} as a baseline. The two non-significant Curve contrasts suggest boundary ambiguity between the Medium and Low tiers, as curved shapes felt similarly rough regardless of speed (see~\cref{tab:ab_summary_main}). To account for this, we defined overlapping candidate pools for haptic--visual matching in \cref{subsec:ps2}: High=\{\emph{sharp\_fast}, \emph{sharp\_slow}, \emph{sharp\_static}\}, Middle=\{\emph{sharp\_slow}, \emph{sharp\_static}, \emph{curve\_fast}\}, Low=\{\emph{curve\_fast}, \emph{curve\_static}, \emph{curve\_slow}\}.}

\replaced{\textbf{Summary.} These results show that contour sharpness was the dominant visual cue for perceived roughness, while motion further modulated roughness in a shape-dependent manner. 
}{Building on the preliminary study, our results confirm that sharp contours are perceived as rougher than curves and that faster motion enhances roughness, although motion differences within curved styles were less reliably distinguished.}

\begin{table}[t]
  \centering
\caption{Visual tiers and candidate pools with corresponding vibrotactile feedback levels.}
  \small
  \begin{tabularx}{\linewidth}{XXXX}
    \toprule
    \textbf{Roughness level} & \textbf{Visual tiers (BT-based)} & \textbf{Visual candidate pool} & \textbf{Vibrotactile feedback} \\
    \midrule
    High & \textit{sharp\_fast}, \textit{sharp\_slow} & \textit{sharp\_fast}, \textit{sharp\_slow}, \textit{sharp\_static} & \textit{I80\_D30} \\
    Middle & \textit{sharp\_static}, \textit{\replaced{curve}{sharp}\_fast} &  \textit{sharp\_slow}, \textit{sharp\_static}, \textit{curve\_fast} & \textit{I80\_D5} \\
    Low    &  \textit{curve\_static}, \textit{curve\_slow} &  \textit{curve\_fast}, \textit{curve\_static}, \textit{curve\_slow} & \textit{I10\_D30} \\
    \bottomrule
  \end{tabularx}
  \label{tab:vh_mapping}
    \vspace{-0.4cm}
\end{table}

\subsection{Preliminary Study~2: Visuo-Haptic Correspondence for Haptic Users}
\label{subsec:ps2}

To address \textbf{RQ1-2}, \replaced{Preliminary Study~2 identified the visual cue that best matched each of the three vibrotactile roughness levels used by the HU in the main study. This validated }{we next validate} a three-level visuo--haptic mapping \replaced{for externalizing }{during touch, so that the cue externalizes} coarse roughness in a form interpretable to a non-haptic partner. \deleted{Specifically, this experiment examined which visual cue best corresponded to each haptic roughness level, and whether participants could interpret roughness consistently when visual and haptic cues were combined.} We therefore treated each visual--haptic pairing as a distinct combined condition rather than comparing visual or haptic cues in isolation.

\textbf{Stimulus.} Vibrotactile feedback used the three levels defined in \cref{sec:tactilefeedback}\replaced{: }{
(}Low=\textit{I10\_D30}, Middle=\textit{I80\_D5}, High=\textit{I80\_D30}\replaced{. These fixed haptic levels were}{),providing fixed haptic conditions without the need for recalibration. 
Each level was} paired with the visual candidate pools \added{from} \cref{subsec:ps1} (see ~\cref{tab:vh_mapping}). Haptic stimuli were designed in bHaptics Designer and delivered \replaced{through}{via} the bHaptics TactGlove DK2 \replaced{via Bluetooth}{(Bluetooth)}. All other software and hardware configurations followed \cref{subsec:ps1}\replaced{, including the visual stimulus designs;}{. Visual stimuli matched the designs used in \cref{subsec:ps1}, and} the overall system setup is illustrated in~\cref{fig:systemsetup}\added{-(b)}.

\begin{table}[t]
\centering
\caption{MatchScore, perceived roughness, and confidence by level and visual. Values are mean (SD). Background colors map roughness scores to levels (Low = green, Middle = blue, High = red).}
\label{tab:match_rough_conf}
\resizebox{\linewidth}{!}{%
\begin{tabular}{llccc}
\toprule
\textbf{Level} & \textbf{Visual} & \textbf{MatchScore} & \textbf{Perceived roughness} & \textbf{Confidence} \\
\midrule
\multirow{3}{*}{High} &
\textit{sharp\_fast}   & \textbf{6.20} (1.15) & \prS{81.96 (10.68)} & 4.43 (0.61) \\
& \textit{sharp\_slow}   & 5.84 (1.22)         & \prS{81.08 (13.13)} & 4.28 (0.80) \\
& \textit{sharp\_static} & 5.61 (1.34)         & \prS{79.22 (11.85)} & 4.14 (0.72) \\
\addlinespace
\multirow{3}{*}{Middle} &
\textit{sharp\_slow}   & \textbf{5.18} (1.41) & \prM{54.51 (17.81)} & 4.14 (0.69) \\
& \textit{curve\_fast}   & 4.82 (1.47)         & \prM{39.71 (19.89)} & 4.04 (0.75) \\
& \textit{sharp\_static} & 4.67 (1.44)         & \prM{49.61 (19.08)} & 4.04 (0.77) \\
\addlinespace
\multirow{3}{*}{Low} &
\textit{curve\_fast}   & \textbf{5.04} (1.34) & \prW{14.31 (7.72)}  & 4.28 (0.75) \\
& \textit{curve\_slow}   & 4.92 (1.61)         & \prW{13.16 (6.43)}  & 4.20 (0.72) \\
& \textit{curve\_static} & 4.69 (1.99)         & \prW{12.59 (7.30)}  & 4.20 (0.75) \\
\bottomrule
\end{tabular}%
}
\end{table}

\textbf{Participants and Procedure.}
Sixteen participants (10 male, 6 female; age range = 22--34 years, $M = 28.2$) took part in the study. All were right-handed \added{and had not participated in \cref{subsec:ps1}}. Most had extensive VR experience (15 had used VR devices more than 10 times), \replaced{while haptic-glove experience was more}{ although only two had fewer than 10 controller-free hand-interaction experiences. In contrast, experience with haptic gloves was} limited: seven had used them fewer than five times, four had used them 5--10 times, and five had used them more than 10 times. \deleted{These participants were a new group and did not take part in \cref{subsec:ps1}.}

Participants completed four blocks\replaced{, each containing}{. Each block contained} the same nine visual--haptic pairings (3 haptic levels $\times$ 3 visual candidates; \cref{tab:vh_mapping})\replaced{ in randomized order.}{, presented once each with randomized trial order within the block.} On each trial, touching a virtual object triggered up to 3~s of vibrotactile feedback\replaced{ with simultaneous visual feedback. A moving bar guided participants}{. The corresponding visual feedback was presented simultaneously, allowing participants to perceive the vibration and visual cue together. Participants were guided by a moving bar at 0.04~m/s} to maintain a consistent hand speed \added{0.04~m/s}, and repetitions were permitted. After each stimulus, \replaced{participants}{they} rated (i) \textit{visual--haptic match} (1 = not at all, 7 = perfectly matches), (ii) \textit{perceived roughness} (1--100, higher = rougher), and (iii) \textit{confidence} (1 = not confident, 5 = very confident). Sessions lasted \deleted{approximately} 20 minutes. The study was approved by the IRB, and participants provided informed consent. Participants received approximately \$8 USD.

\textbf{Results.} \replaced{Because the visual candidates differed by haptic level, we treated the nine visual--haptic pairings as a single within-subject factor and applied Friedman tests. Condition significantly affected }{ As the stimuli were not fully crossed, the nine conditions were treated as a single within-subject factor. Because the data for \textit{visual–match score}, \textit{perceived roughness}, and \textit{confidence} 
violated normality assumptions, we applied Friedman tests for repeated measures. Results showed significant effects of condition on}\textit{MatchScore} 
($\chi^2(8)=35.20$, $p<.001$, $W=0.28$) and \textit{Perceived Roughness} 
($\chi^2(8)=120.00$, $p<.001$, $W=0.94$), but not \deleted{on} \textit{Confidence} 
($\chi^2(8)=12.70$, $p=.122$, $W=0.10$).

\replaced{As summarized in \cref{tab:match_rough_conf}, the highest \textit{MatchScore} for each haptic level was obtained by \textit{sharp\_fast} for High, \textit{sharp\_slow} for Middle, and \textit{curve\_fast} for Low. Perceived roughness ratings also followed the intended order, increasing monotonically from Low ($M\approx13$) to Middle ($M\approx45$) and High ($M\approx80$). Confidence remained moderate to high across conditions ($M=4.0$--$4.4$/5), supporting the consistency of these judgments.}{ We then examined descriptive statistics to interpret patterns in visual–haptic matching. 
To identify the best-matching visual stimulus for each haptic level, 
we compared mean \textit{MatchScore} values across the three candidate visuals (see~\cref{tab:match_rough_conf}). 
The highest-scoring visual was considered the best match. 
For the \emph{High} level, \textit{sharp\_fast} yielded the highest score ($M=6.20, SD=1.15$). 
At the \emph{Middle} level, \textit{sharp\_slow} aligned best ($M=5.18, SD=1.41$), 
and for the \emph{Low} level, \textit{curve\_fast} was rated highest ($M=5.04, SD=1.34$). As a validity check, we examined whether perceived roughness ratings 
followed the intended order of roughness levels 
(\textit{Low} $<$ \textit{Middle} $<$ \textit{High}). 
Mean ratings increased monotonically (Low $\approx$ 13, Middle $\approx$ 45, High $\approx$ 80), 
indicating that participants consistently linked visual differences with  tactile roughness level.
Confidence ratings remained moderate to high across conditions ($M=4.0$–$4.4$ on a 5-point scale), 
supporting the reliability of these judgments.}

Post-task interviews \replaced{suggested}{revealed} that participants judged visual--haptic matching 
primarily by \replaced{edge sharpness, motion speed, and temporal congruence between vibration and visual motion.}{\emph{edge sharpness} (Sharp vs.\ Curve), \emph{motion speed} (Static/Slow/Fast), 
and \emph{temporal congruence} between vibration and visual motion. }\replaced{ Participants associated sharper and faster-moving cues with higher roughness; for example, P8 noted that the surface felt rougher when the visual cue ``moved quickly or had sharper edges.'' Static variants scored lower because they felt less aligned with rhythmic vibration (P1, P10). For Middle, \textit{curve\_fast} offered a suitable speed cue but appeared too soft, explaining its lower match score than \textit{sharp\_slow}.}{
Faster motion and sharper edges were generally perceived as rougher 
(e.g., ``When the visual moved quickly or had sharper edges, the surface felt rougher to me'' (P8)). 
Static variants scored lowest across all haptic levels (see~\cref{tab:match_rough_conf}), 
as they felt poorly aligned with rhythmic vibration 
(e.g., ``a static visual did not seem to match'' (P1); 
``animated motion fit when aligned with vibration frequency; when misaligned, the mismatch was conspicuous'' (P10)). 
At the Middle level, participants noted that although \textit{curve\_fast} provided a suitable speed cue, 
its curved (softer) shape conflicted with the expected slight sharpness, 
yielding lower ratings (P10, P11).}

Following \replaced{the selection rule based on \textit{MatchScore} and roughness ratings,}{the preregistered rule (highest \textit{MatchScore} consistent with roughness ratings) and informed by interview insights,} we selected \textit{sharp\_fast} (High), \textit{sharp\_slow} (Middle), and \textit{curve\_fast} (Low) as the final mappings (see~\cref{tab:visuo_haptic_mapping}). \added{Together, these results established the three-level visuo--haptic roughness mapping used to externalize the HU's tactile state in the main study.}

\begin{table}[t]
  \centering
  \small
  \caption{Final visuo--haptic mapping used in the main study.}
  \label{tab:visuo_haptic_mapping}
  \begin{tabular}{@{}lll@{}}
    \toprule
    \textbf{Roughness level} & \textbf{Visual style} & \textbf{Haptic pattern} \\
    \midrule
    High   & \texttt{sharp\_fast} & \texttt{I80\_D30} \\
    Middle & \texttt{sharp\_slow} & \texttt{I80\_D5}  \\
    Low    & \texttt{curve\_fast} & \texttt{I10\_D30} \\
    \bottomrule
  \end{tabular}
    \vspace{-0.4cm}
\end{table}

\subsection{Summary of Findings}

\replaced{Together, the preliminary studies establish a compact visual--haptic mapping for externalizing coarse roughness in the main collaborative task, rather than validating realistic haptic rendering.} {Overall, the preliminary studies validate \textit{Line shape} and \textit{Motion} as effective dimensions for externalizing coarse tactile roughness. We draw three key takeaways.}

\textbf{(1) Visual discriminability for non-haptic users (RQ1-1).}
Non-haptic users reliably differentiated the cue variants: sharp contours were \deleted{consistently} judged rougher than curved contours, while motion \deleted{further} modulated perceived roughness in a shape-dependent manner. \deleted{This supports using contour sharpness and motion as glanceable visual signals of roughness.}

\textbf{(2) Visuo--haptic correspondence for haptic users (RQ1-2).}
Haptic users \replaced{aligned the visual candidates with the three vibrotactile roughness levels, with contour sharpness serving as the primary cue and motion helping differentiate selected mappings.}{consistently mapped the same visual dimensions to tactile roughness, aligning higher roughness with sharper contours and faster motion.} Roughness judgments followed the intended ordering, indicating that the cue can externalize coarse roughness in an interpretable form.

\textbf{(3) Parameter selection for the main visualization.}
Across \replaced{the preliminary studies,}{both roles,} contour sharpness emerged as the dominant cue and motion provided a secondary cue. We therefore selected shape--motion combinations that yielded separable \replaced{Low/Middle/High mappings while accounting for weaker separability within the curved family.}{low/middle/high tiers and avoid ambiguous pairings observed in the curved family.}

With these prerequisites established, the main study (\cref{subsec:4}) evaluates \replaced{who receives the externalized tactile-state cue by toggling whether it is rendered on the HU's and/or NU's hand.}{a collaboration question: when tactile state is externalized, who should receive the cue? We isolate recipient visibility by independently toggling whether the cue is rendered on the HU’s and/or NU’s hand.}

\section{Evaluating Cue Visibility in Asymmetric Collaboration}
\label{subsec:4}

\begin{table}[t]
\centering
\scriptsize
\setlength{\tabcolsep}{3pt}
\caption{What each role experiences on their own hand. We toggle cue rendering on the HU’s and/or NU’s hand (self-anchored view); both can also observe the partner’s hand.}
\label{tab:role_placement}
\begin{tabular}{lcc}
\toprule
\textbf{Condition} & \textbf{HU’s hand} & \textbf{NU’s hand} \\
\midrule
\textit{Off}  & Haptic only & None \\
\textit{H}    & Haptic + visual cue & None \\
\textit{N}    & Haptic only & Visual cue only \\
\textit{Both} & Haptic + visual cue & Visual cue only \\
\bottomrule
\end{tabular}
\vspace{-0.2cm}
\end{table}

Having established a shared visual vocabulary for roughness, we next examine who should receive the externalized tactile cue in asymmetric collaboration. Specifically, we ask whether collaboration benefits arise from reducing the non-haptic user’s (NU) information disadvantage by showing the haptic user’s (HU) tactile-state cue on the NU’s hand, and whether additionally showing the cue on the HU’s hand provides further benefits through self-feedback.

To test this, we use a $2\times2$ placement design that \deleted{independently }toggles whether the hand-anchored cue encoding the HU’s \deleted{current }tactile state is shown on the HU’s hand and/or the NU’s hand. This yields four conditions: \textit{Off} (neither), \textit{H} (HU only), \textit{N} (NU only), and \textit{Both} (both). Here, \textit{Off} serves as a no-cue baseline, \textit{N} tests NU-side cue placement, \textit{H} isolates HU-side self-feedback, and \textit{Both} tests whether shared placement provides benefits beyond NU-only placement.

\noindent\textbf{RQ2.} In asymmetric haptic collaboration, how does placing a hand-anchored visual proxy for an object’s tactile state on the non-haptic user’s and/or haptic user’s hand influence task performance, collaboration quality, and social presence?

\subsection{Independent Variables}
We manipulated \textit{Role} and \textit{Mode} as independent variables (see~\cref{tab:role_placement}). The collaboration setting was \textit{asymmetric--haptic}: in each pair, one participant wore a vibrotactile glove (\textit{haptic user}, HU) while the other did not (\textit{non-haptic user}, NU). Thus, \textit{Role} had two levels: \textit{HU} and \textit{NU} (see~\cref{fig:teaser}-(b)).

\textit{Mode} specifies where the hand-outline visualization is rendered during object contact. Although both collaborators can observe each other’s hands, our manipulation determines whether the cue is rendered on the HU’s hand and/or the NU’s hand (i.e., first-person, self-anchored visibility). This yields four modes (see~\cref{fig:teaser}-(c)): \textit{Off} (no visualization), \textit{H} (visualization on HU only), \textit{N} (visualization on NU only), and \textit{Both} (visualization on both users). Feedback was triggered only when a hand touched a cube and reflected the cube’s assigned roughness level. The HU received vibrotactile feedback during contact, and the hand-outline visualization (when enabled) was rendered accordingly.

\begin{table}[t]
\centering
\caption{Participant summary (N=48; 24 HU--NU pairs).}
\label{tab:participants_summary}
\small
\setlength{\tabcolsep}{4pt}
\renewcommand{\arraystretch}{0.95}
\begin{tabular}{@{}p{0.24\columnwidth}p{0.70\columnwidth}@{}}
\toprule
\textbf{Measure} & \textbf{Details} \\
\midrule
\rowcolor{gray!12}\multicolumn{2}{@{}l@{}}{\textbf{Demographics and pairing}} \\
Participants & 48 (23 female, 25 male) \\
Age (years) & 20--35 ($M=27.4$, $SD=3.3$) \\
Assignment & 24 pairs (HU--NU), random pairing \\
Acquaintance & None (all pairs unacquainted) \\
Pair genders & 9 F--M (37.5\%), 8 M--M (33.3\%), 7 F--F (29.1\%) \\
Handedness & 2 left-handed (HU); all others right-handed \\
\addlinespace
\rowcolor{gray!12}\multicolumn{2}{@{}l@{}}{\textbf{Prior experience (counts out of 24 per role)}} \\
VR/AR & Low/Mid/High: HU 9/4/11; NU 11/4/9 \\
Bare-hand & Low/Mid/High: HU 15/3/6; NU 19/0/5 \\
Haptic glove & Low/Mid/High: HU 22/0/2 \\
\bottomrule
\end{tabular}
\footnotesize \textit{Note.} Low=0--4 uses; Mid=5--10; High$>$10.
\vspace{-0.6cm}
\end{table}

\subsection{Study Design and Procedure}

We fixed the visuo--haptic mapping (see~\cref{tab:visuo_haptic_mapping}) and studied 24 asymmetric-haptic HU--NU dyads (48 participants) performing a collaborative roughness-sorting task. Demographics and prior experience are summarized in \cref{tab:participants_summary}. The study was approved by an IRB. Participants provided informed consent and completed a demographic questionnaire\replaced{. Before the main task, the experimenter performed calibration checks for headset fit, hand tracking, network connection, and HU vibrotactile output. Participants then }{, and}practiced cube grabbing and three-level roughness discrimination with visuo--haptic feedback. \added{The study was conducted in a shared physical room of approximately 8.04\,m $\times$ 6.70\,m. The HU and NU stood in separate, non-overlapping stationary play areas of approximately 2.68\,m $\times$ 2.23\,m, positioned diagonally apart and separated by a physical partition for safety. Dyads communicated using natural voice in the shared room.}

Each dyad then \deleted{stood at assigned positions, }greeted each other in VR\deleted{,} and completed the main task under four visualization \textit{Mode} conditions. To reduce practice and fatigue effects, we counterbalanced the order of the four \textit{Mode} blocks across dyads using a Latin-square schedule. Each \textit{Mode} was administered as a block with two trials. In each trial, nine visually identical cubes were randomly assigned to three roughness levels with either a 1--4--4 or 3--4--2 distribution (Low/Middle/High). After pressing \textit{Start}, the dyad collaboratively sorted the cubes into the three labeled zones and pressed \textit{Finish} when satisfied. After each \textit{Mode} block, participants completed questionnaires and took a short break before proceeding. The task had no time limit, and dyads were encouraged to move and communicate freely. After completing \deleted{all} four modes, participants were interviewed separately to rank the modes and reflect on their collaborative experience, avoiding mutual influence between partners. Each session lasted \deleted{approximately} 75 minutes, and participants received \$15 USD.

\subsection{Implementation and Setup}
To support multiuser interaction, the system was \replaced{implemented in Unity 6000.1.6f1 using Netcode for GameObjects (NGO), which synchronized shared object states, interaction events, and outline-cue visibility across clients. Because the outline shader \replaced{differed}{appeared differently} between standalone Quest and PC execution, \deleted{the} two Meta Quest 3 headsets were connected to separate PCs via Meta Quest Air Link. Both PCs were on the same local network for synchronized interaction and experimenter monitoring, and each had 64\,GB RAM and an RTX 4090-class GPU.}{configured by adapting the Unity VR Multiplayer Template, which provides room creation, object synchronization, and voice chat. Synchronization of the outline visualization was additionally implemented via Unity networking.}

\begin{figure}[t]
  \centering
  \includegraphics[width=\columnwidth]{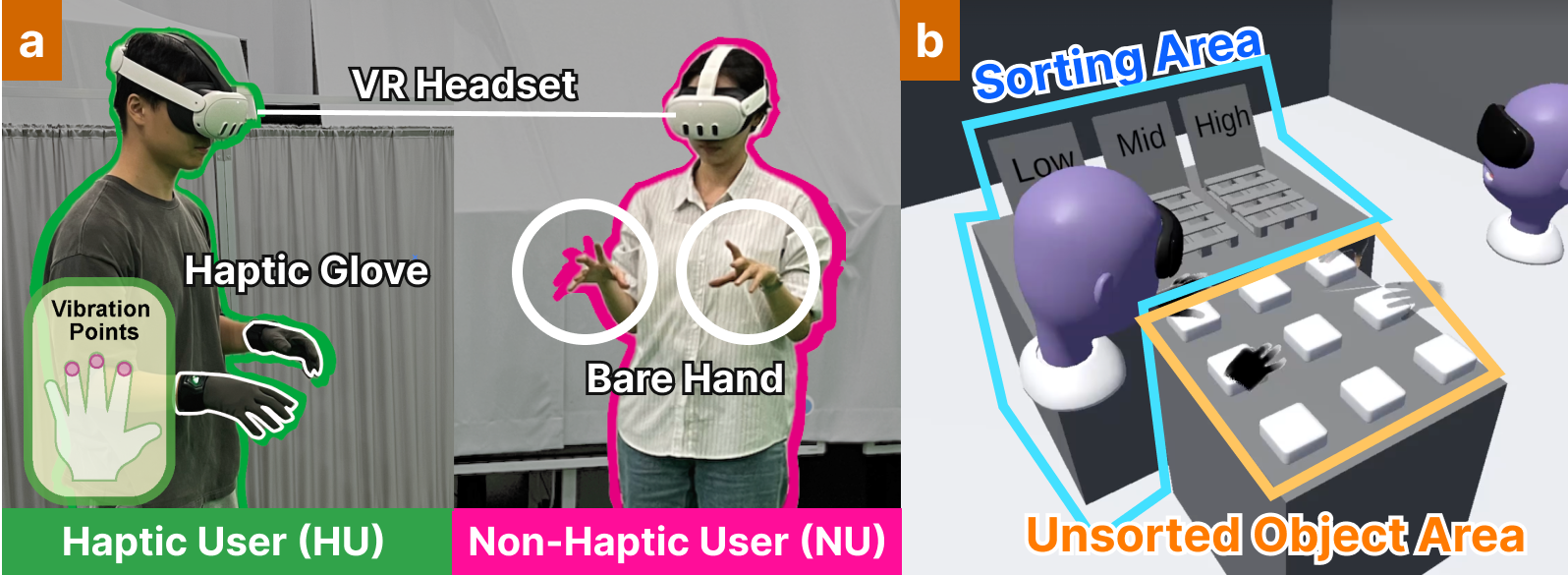}
    \caption{Main study setup and sorting task. (a) Setup of the haptic user (HU) and non-haptic user (NU). (b) Sorting workspace with nine cubes, an unsorted object area, and Low/Mid/High sorting areas. Examples of cue placement on the HU's and NU's virtual hands are shown in \cref{fig:teaser}(c).} 
  \label{fig:setupmain}
\end{figure}

As in the preliminary studies, stimuli were delivered through the bHaptics TactGlove DK2, and the VR environment was built in the same Unity version. \added{The main study used three HU vibrotactile patterns: Low=\textit{I10\_D30}, Middle=\textit{I80\_D5}, and High=\textit{I80\_D30}. I10/I80 denote normalized bHaptics intensity settings, not frequencies, and D5/D30 denote 5 ms and 30 ms vibration on-durations. Feedback targeted the index, middle, and ring fingertips using the TactGlove DK2.} 
\deleted{Two Meta Quest 3 headsets were used, with both instances executed on PCs on the same local network to address discrepancies in outline shader behavior between standalone and PC builds and to enable synchronized interaction on a shared server. Although the Unity template supported voice chat, this feature was disabled due to latency.} \replaced{\cref{fig:setupmain} shows the main study setup and sorting workspace. The HU wore a vibrotactile glove, whereas the NU interacted with bare hands. Dyads communicated by voice in the shared room. In the sorting workspace, dyads classified nine visually identical cubes from the unsorted object area into Low/Mid/High zones.}{~\cref{fig:setupmain} shows the experimental setup with HU wearing a vibrotactile glove and NU using bare hands.}

\deleted{The study was conducted in a shared physical room to allow natural voice communication. 
Although the Unity template supported voice chat, this feature was disabled due to latency issues. 
To prevent unintended physical collision and ensure safety, a partition was placed between participants, 
and the stationary boundary system was employed to restrict movement to predefined play areas.}

\begin{table}[t]
 \centering
    \caption{Questionnaires for collaboration rating (10 items across 4 subscales).}
 \includegraphics[width=\columnwidth]{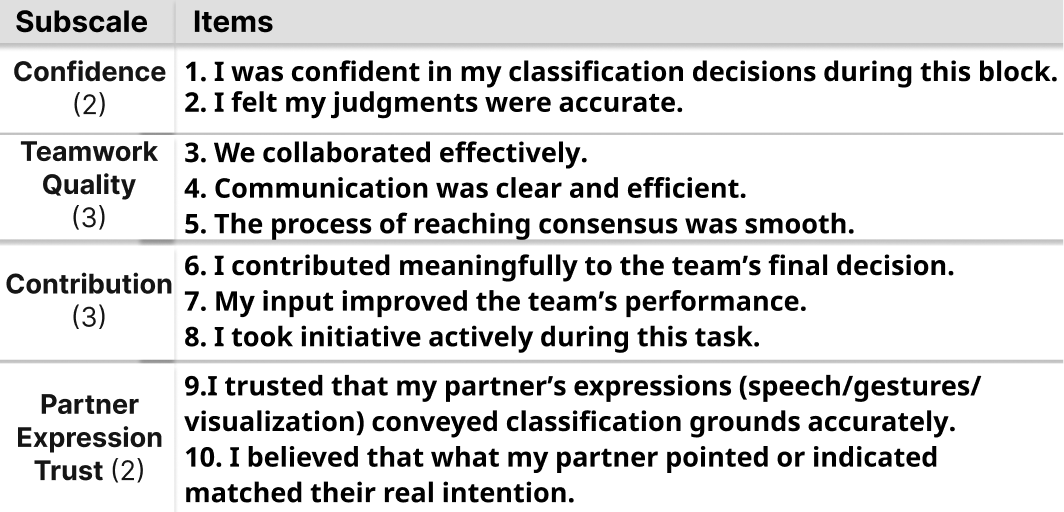} 
 \label{tab:collabo}
 \vspace{-0.3cm}
\end{table}

\subsection{Dependent Variables}
We evaluated the following dependent variables. All Likert-type items used a 7-point scale (1 = strongly disagree, 7 = strongly agree). (i) \textbf{Task performance}: total \replaced{completion}{task} time and success rate. \added{Success rate was computed for each trial as the mean of the correct-classification rates for the Low, Middle, and High roughness levels.}
(ii) \textbf{Interaction measures}: total \replaced{cube grasp}{grab} time per \textit{Role}.  
(iii) \textbf{Social presence}: Networked Minds~\cite{harms2004internal}, using 24 adapted items across four subscales (co-presence, attentional allocation, perceived message understanding, perceived behavioral interdependence).  
(iv) \textbf{Perceived workload}: NASA--TLX~\cite{hart1988development}\added{, reported as raw overall workload (RTLX) and subscale scores.} 
(v) \textbf{Collaboration ratings}: 10 custom items across four dimensions (Confidence, Teamwork Quality, Contribution, Partner Expression Trust), adapted from established teamwork measures (see~\cref{tab:collabo}).  
(vi) \textbf{Post-study interview}: participants ranked the four visualization \textit{Mode} conditions and reflected on their collaborative experience. Interviews were conducted separately to avoid mutual influence between partners.

\begin{figure}[t]
  \centering
  \includegraphics[width=\columnwidth]{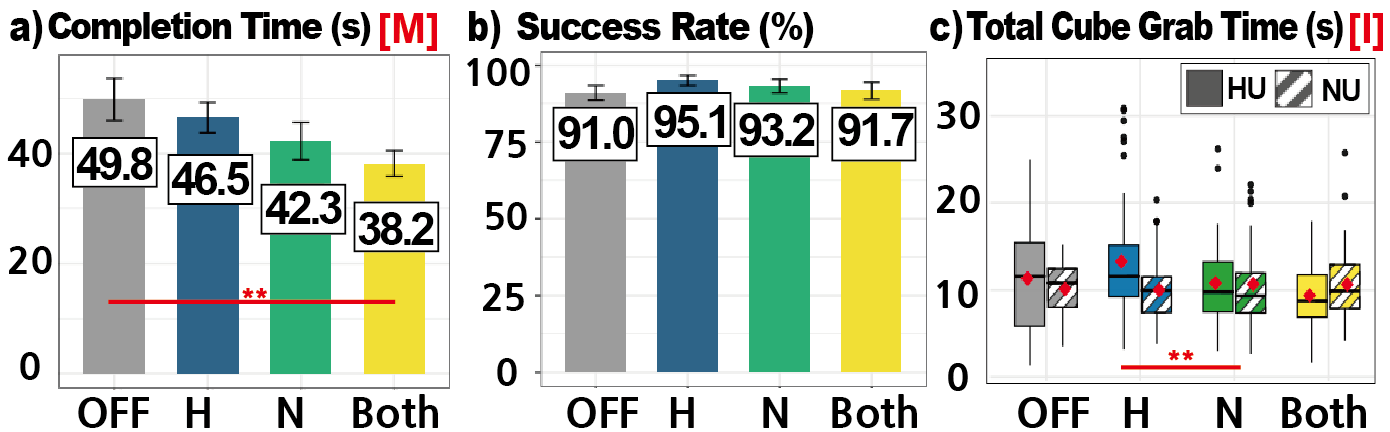}
  \caption{Task performance results. (a) Completion time. (b) Success rate, defined as the mean correct-classification rate across the three roughness levels. (c) Total cube grasp time. (Red M: significant main effect of \textit{Mode}; I: significant interaction effect) 
}
  \label{fig:flowchart2}
\vspace{-0.5cm}
\end{figure}

\begin{table*}[t]
\centering
\caption{Summary of main-study inferential statistics.
\textbf{M} = main effect of \textit{Mode};
\textbf{R} = main effect of \textit{Role};
\textbf{I} = \textit{Mode} $\times$ \textit{Role} interaction.
Only significant effects and key post-hoc patterns are shown;
non-significant effects are omitted unless central to interpretation.}
\label{tab:main_stats_summary}

\footnotesize
\fontsize{7.5pt}{8.5pt}\selectfont
\setlength{\tabcolsep}{2.5pt}
\renewcommand{\arraystretch}{1.02}

\begin{tabularx}{\textwidth}{
@{}
>{\raggedright\arraybackslash}p{0.17\textwidth}
>{\raggedright\arraybackslash}p{0.53\textwidth}
>{\raggedright\arraybackslash}X
@{}
}
\toprule
\textbf{Measure}
& \textbf{Significant effects}
& \textbf{Main pattern} \\
\midrule

Completion time
& \textbf{M}: $\chi^2(3)=14.6$, $p=.002$, $W=.20$
& Both $<$ Off ($p=.006$) \\

Success rate
& n.s.
& Accuracy remained high \\

Total cube grasp time
& \textbf{I}: $F(3,346)=2.82$, $p=.039$, $\eta^2_p=.024$
& H/N role reversal ($p=.004$) \\

\midrule

Overall workload
& \textbf{M}: $F(3,126)=3.35$, $p=.021$, $\eta^2_p=.074$;
  \textbf{I}: $F(3,126)=5.12$, $p=.002$, $\eta^2_p=.109$
& Both $<$ H; largest HU--NU gap in Off \\

Temporal demand
& \textbf{M}: $F(3,126)=4.79$, $p=.003$, $\eta^2_p=.10$
& H $>$ Both, N \\

Performance
& \textbf{M}: $F(3,126)=4.03$, $p=.009$, $\eta^2_p=.09$
& Both $>$ Off \\

Effort
& \textbf{I}: $F(3,126)=3.58$, $p=.016$, $\eta^2_p=.08$
& Off: HU $>$ NU \\

Mental / Physical / Frustration
& n.s.
& No reliable effects \\

\midrule

Confidence
& \textbf{M}: $F(3,138)=13.59$, $p<.001$, $\eta^2_p=.23$;
  \textbf{R}: $F(1,46)=15.82$, $p<.001$, $\eta^2_p=.26$;
  \textbf{I}: $F(3,138)=10.48$, $p<.001$, $\eta^2_p=.19$
& Off $<$ N, Both; NU confidence highest in N/Both \\

Contribution
& \textbf{M}: $F(3,138)=10.15$, $p<.001$, $\eta^2_p=.18$;
  \textbf{R}: $F(1,46)=24.47$, $p<.001$, $\eta^2_p=.35$;
  \textbf{I}: $F(3,138)=31.32$, $p<.001$, $\eta^2_p=.41$
& Off $<$ N, Both; NU contribution highest in N/Both \\

Partner expression trust
& \textbf{R}: $F(1,46)=12.49$, $p=.001$;
  \textbf{I}: $F(3,138)=3.19$, $p=.026$
& NU $>$ HU; no within-role mode differences \\

Teamwork quality
& n.s.
& No reliable effects \\

Social presence
& n.s.
& Stable across conditions \\

\bottomrule
\end{tabularx}

\vspace{-0.3cm}
\end{table*}

\begin{figure*}[t]
\centering       \includegraphics[width=0.95\textwidth]{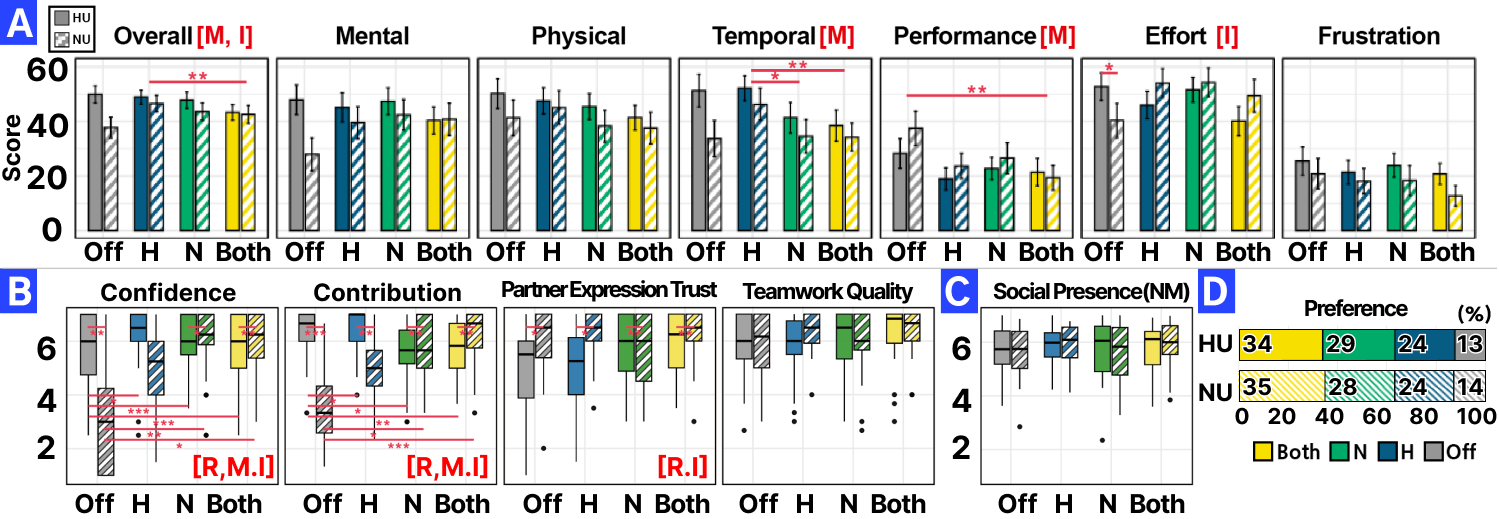}
\caption{Results for (a) overall workload (RTLX) and NASA-TLX subscales, (b) collaboration ratings, (c) social presence (NM) and (d) preference. (Red R and M: significant main effect of \textit{Mode} and \textit{Role}; I: significant interaction effect)} 
    \label{fig:graph_all}
\end{figure*}

\section{RESULTS}
 We analyzed the data using RStudio (version 4.4.1). Normality was assessed with the Shapiro–Wilk test. 
For objective measures (Completion Time, Success Rate), we treated \textit{Mode} (\textit{Off}, \textit{H}, \textit{N}, \textit{Both}) as a within-subject factor and conducted one-way repeated-measures ANOVAs. 
When \replaced{normality}{the normality assumption} was violated, we applied the non-parametric Friedman test instead. 
Post-hoc comparisons \deleted{pairwise comparisons }were performed using paired-samples $t$-tests (or Wilcoxon signed-rank tests when normality was violated), with Bonferroni-adjusted $p$-values. As violations of normality were frequent, we report only non-parametric results.

For subjective measures and Total Cube Grasp Time, we used a mixed factorial design with \textit{Mode} (within-subjects) and \textit{Role} (between-subjects: HU = Haptic User, NU = Non-haptic User). 
\replaced{Data}{These data} were analyzed with the Aligned Rank Transform (ART) for factorial nonparametric analysis ($\alpha = .05$) following Wobbrock et al.~\cite{wobbrock2011aligned}. 
\replaced{Post-hoc comparisons used }{Post-hoc analyses were conducted using }\texttt{emmeans} pairwise comparisons with Tukey correction. 
Effect sizes were reported as Kendall’s $W$ for Friedman tests, $r$ for Wilcoxon signed-rank tests, partial eta squared ($\eta^2_p$) for ART ANOVAs, and Cohen’s $d$ for post-hoc comparisons. 
Internal consistency \deleted{of multi-item scales }was assessed using Cronbach’s $\alpha$. \added{Inferential statistics are summarized in \cref{tab:main_stats_summary}. Below, we focus on significant patterns and interpretation.}

\subsection{Objective \replaced{Measures}{Ratings}}
\added{Task performance results are shown in \cref{fig:flowchart2}, and inferential statistics summarized in \cref{tab:main_stats_summary}.} \added{Overall, the visualization improved sorting efficiency without reducing accuracy, while cue placement shifted which partner handled and interpreted tactile-state evidence.}

\textbf{Completion Time \added{and Success Rate.}} \replaced{Completion time differed by \textit{Mode}, with \textit{Both} ($M=38.19$\,s, $SD=11.57$) significantly faster than \textit{Off} ($M=49.83$\,s, $SD=18.75$). In contrast, success rate did not differ across modes and remained high ($M=91.0$--$95.1\%$). These results indicate that visualization improved task efficiency without reducing sorting accuracy.}{ The results of task performance is shown in \cref{fig:flowchart2}. Completion time was significantly affected  \textit{Mode}, as indicated by a Friedman test, $\chi^2(3)=14.6$, $p=.002$, $W=0.20$. Post-hoc Wilcoxon signed-rank tests with Bonferroni correction revealed that the \textit{Both} condition ($M=38.19$ s, $SD=11.57$) was significantly faster than \textit{Off} ($M=49.83$ s, $SD=18.75$, $p=.006$). No other pairwise comparisons were significant. }\textbf{\deleted{Success Rate.}} \deleted{Success rate was not significantly affected by \textit{Mode}, Friedman test, $\chi^2(3)=3.40$, $p=.334$. These findings suggest that visualization improved task efficiency (completion time), while accuracy (success rate) remained uniformly high across conditions.}

\textbf{Total Cube Grasp Time per Trial.} \replaced{Total Cube Grasp Time showed only a significant \textit{Mode} $\times$ \textit{Role} interaction, with no main effects of either factor. The role pattern reversed between \textit{H} and \textit{N}: grasp time was longer for the HU in \textit{H} (HU/NU: $13.27(6.96)/10.01(3.58)$\,s), but longer for the NU in \textit{N} (HU/NU: $9.38(3.50)/10.63(4.04)$\,s; values are $M(SD)$). This suggests that cue placement shifted which partner more actively handled and interpreted the cubes.}{Most conditions met the normality assumption, with one exception; non-parametric tests were used. Total Cube Grasp Time revealed no significant main effects of \textit{Mode}($F(3,346)=1.47$, $p=.222$, $\eta^2_p=.013$), or \textit{Role}($F(1,347)=1.77$, $p=.184$, $\eta^2_p=.005$).
However, the \textit{Mode} $\times$ \textit{Role} interaction was significant, 
$F(3,346)=2.82$, $p=.039$.$\eta^2_p=.024$.
Post-hoc pairwise comparisons showed that the difference between \textit{H} (HU vs.\ NU) and \textit{N} (HU vs.\ NU) was significant ($p=.004$), whereas other contrasts were not. 
In \textit{H}, the mean grasp time was longer when the HU held the cube ($M=13.27$\,s, $SD=6.96$) than when the NU held it ($M=10.01$\,s, $SD=3.58$). 
In contrast, in \textit{N}, the mean grasp time was longer when the NU held the cube ($M=10.63$\,s, $SD=4.04$) than when the HU held it ($M=9.38$\,s, $SD=3.50$). 
Thus, the direction of the Role effect differed across conditions, driving the significant interaction.}

\subsection{Subjective \replaced{Measures}{Ratings}}
\subsubsection{Task Load}

\added{Task-load results suggest that tactile-state visualization reduced workload imbalance and improved perceived task performance without increasing other workload dimensions.}

\textbf{Overall workload.} \replaced{Overall workload showed a significant effect of \textit{Mode} and a significant \textit{Mode} $\times$ \textit{Role} interaction (see~\cref{tab:main_stats_summary}). Post-hoc tests for the \textit{Mode} effect showed that workload was lower in \textit{Both} than in \textit{H} ($p=.045$; HU/NU mean scores: $43.27/42.55$ vs.\ $48.86/46.54$). For the interaction, the HU--NU workload gap was largest in \textit{Off}, where HUs reported higher workload than NUs ($M=49.88$, $SD=14.63$ vs.\ $M=37.77$, $SD=17.64$); this gap was significantly greater than in \textit{Both}, \textit{H}, and \textit{N} (all $p\leq.020$; see~\cref{fig:graph_all}-(a)). No other post-hoc comparisons were significant.}{
For overall workload (RTLX), there was a significant main effect of \textit{Mode} ($F(3,126)=3.35$, $p=.021$, $\eta^2_p=.074$), but no significant main effect of \textit{Role}. There was also a significant \textit{Role} \(\times\) \textit{Mode} interaction ($F(3,126)=5.12$, $p=.002$, $\eta^2_p=.109$). For the \textit{Mode} effect, post-hoc Tukey tests averaged over \textit{Role} showed that workload in the \textit{Both} condition was significantly lower than in the \textit{H} condition ($p=.045$; HU/NU means: 43.27/42.55 vs.\ 48.86/46.54). No other pairwise differences reached significance ($p \geq .26$). For the interaction effect, post-hoc comparisons showed that the HU--NU difference was largest in the \textit{Off} condition, with HU reporting higher workload ($M=49.88$, $SD=14.63$) than NU ($M=37.77$, $SD=17.64$). This gap was significantly greater than in \textit{Both}, \textit{H}, and \textit{N} ($p<.001$, $p<.001$, and $p=.020$, respectively). In the other conditions, HU and NU workload ratings were much closer (\textit{Both}: 43.27 vs.\ 42.55; \textit{H}: 48.86 vs.\ 46.54; \textit{N}: 47.80 vs.\ 43.53), and no significant HU--NU differences were found ($p>.30$; see~\cref{fig:graph_all}-(a)).}

\textbf{\added{NASA--TLX Subscales.}} \added{Temporal Demand and Performance showed significant effects of \textit{Mode}. For Temporal Demand, post-hoc tests showed that \textit{H} was higher than \textit{Both} ($p=.004$) and \textit{N} ($p=.015$); descriptively, \textit{H} showed higher HU/NU mean scores ($52.14/46.23$) than \textit{Both} ($38.45/34.23$) and \textit{N} ($41.36/34.55$). For Performance, \textit{Both} was rated significantly higher than \textit{Off} ($p=.007$; HU/NU mean scores: $78.64/80.64$ vs.\ $71.77/62.55$), while \textit{H} showed a marginally higher score than \textit{Off} ($p=.050$; $81.05/76.36$ vs.\ $71.77/62.55$). Effort showed \deleted{only }a significant \textit{Mode} $\times$ \textit{Role} interaction, with no main effects of either factor: HUs reported higher effort than NUs only in \textit{Off} ($M=52.77$, $SD=23.73$ vs.\ $M=40.32$, $SD=29.39$, $p\leq.036$). Mental Demand, Physical Demand, and Frustration showed no reliable effects. }
\textbf{\deleted{Temporal Demand.}} \deleted{For Temporal Demand, there was a significant main effect of \textit{Mode} ($F(3,126)=4.79$, $p=.003$, $\eta^2_p=.10$), while no main effect of \textit{Role} or interaction was observed. Post-hoc Tukey tests (averaged over \textit{Role}) confirmed that the \textit{H} condition elicited significantly higher temporal demand than both the \textit{Both} ($p=.004$) and \textit{N} conditions ($p=.015$). Descriptively, \textit{H} showed higher HU/NU means (52.14/46.23) than \textit{Both} (38.45/34.23) and \textit{N} (41.36/34.55). Although \textit{Off} also showed lower values than \textit{H} (51.23/33.68), this difference did not reach significance ($p=.28$).} \textbf{\deleted{Performance.}} \deleted{For Performance, there was a significant main effect of \textit{Mode} ($F(3,126)=4.03$, $p=.009$, $\eta^2_p=.09$), while no main effect of \textit{Role} or interaction was observed. Post-hoc Tukey tests revealed that performance in the \textit{Both} condition was significantly higher than in \textit{Off} ($p=.007$; HU/NU means: 78.64/80.64 vs.\ 71.77/62.55). The \textit{H} condition also showed a marginally higher score than \textit{Off} ($p=.050$; 81.05/76.36 vs.\ 71.77/62.55). No other pairwise differences reached significance ($p \geq .24$).} \textbf{\deleted{Effort.}} \deleted{For Effort, there was a significant \textit{Role} \(\times\) \textit{Mode} interaction ($F(3,126)=3.58$, $p=.016$, $\eta^2_p=.08$), while neither the main effect of \textit{Role} nor \textit{Mode} was significant. Post-hoc comparisons revealed that HU reported significantly higher effort than NU in the \textit{Off} condition ($M=52.77$, $SD=23.73$ vs.\ $M=40.32$, $SD=29.39$; $p \leq .036$), whereas no HU--NU differences were observed in the \textit{Both}, \textit{H}, or \textit{N} conditions (all $p \geq .48$).
For Mental Demand, no significant main effects or interaction were found (all $p \geq .05$). For Frustration, neither the main effects of \textit{Role} and \textit{Mode} nor interaction reached significance (all $p \geq .16$).}

\subsubsection{Collaboration Ratings}
\added{Collaboration ratings indicate that cue placement affected confidence and contribution, especially for NUs, rather than general teamwork quality.} The composite scale showed acceptable reliability (Cronbach's $\alpha=.80$)\replaced{, with subscale reliabilities ranging from acceptable to excellent ($\alpha=.71$--.95). Results are shown in \cref{fig:graph_all}-(b), and inferential statistics are summarized in \cref{tab:main_stats_summary}.}{The composite collaboration rating showed excellent reliability (Cronbach’s $\alpha = .80$). 
Subscales were also acceptable to excellent: Confidence ($\alpha = .95$), Teamwork Quality ($\alpha = .86$), Contribution ($\alpha = .82$), and Partner Expression Trust ($\alpha = .71$). As shown in \cref{fig:graph_all}-(b), the details of statistical results are provided.}

\textbf{Confidence \added{and Contribution}.}
Confidence and Contribution showed significant effects of
\textit{Mode}, \textit{Role}, and their interaction. For confidence,
ratings averaged across roles were lower in \textit{Off} than in
\textit{N} and \textit{Both}
(see \cref{tab:main_stats_summary}). HUs maintained relatively high
confidence across modes ($M=5.52$--$6.15$), whereas NU confidence
varied more strongly with cue placement: it was lowest in \textit{Off}
($M=3.23$), higher in \textit{H} ($M=4.88$), and highest in
\textit{N} ($M=5.90$) and \textit{Both} ($M=5.96$).

Similarly, contribution ratings averaged across roles were lower in
\textit{Off} than in \textit{N} and \textit{Both}. NU contribution
was lowest in \textit{Off} ($M=3.51$), higher in \textit{H}
($M=4.92$), and highest in \textit{N} ($M=5.79$) and
\textit{Both} ($M=6.13$). Mean contribution was higher for HUs in
\textit{Off} and \textit{H}, but higher for NUs in \textit{N} and
\textit{Both}. These patterns suggest that NU-side cue visibility
narrowed role differences in confidence and shifted perceived
contribution toward NUs.

\textbf{Partner Expression Trust\added{ and Teamwork Quality}.} Partner Expression Trust showed a significant effect of \textit{Role} and a significant \textit{Mode} $\times$ \textit{Role} interaction (see~\cref{tab:main_stats_summary}). NUs reported higher trust than HUs overall, and this role difference was significant in all four \textit{Mode} conditions (all $p<.05$); however, neither role \replaced{differed}{showed reliable differences} across modes. Teamwork Quality showed no reliable effects. \textbf{\deleted{Teamwork Quality.}}\deleted{For Teamwork Quality, no significant main effects or interaction were found (all $p \geq .12$).}

\subsubsection{Social Presence}
\added{Social presence remained stable across cue placements, while preferences favored \textit{Both} and highlighted NU-side cue visibility.}

\textbf{Networked Minds Social Presence (NM).} The composite NM score showed excellent reliability (Cronbach’s $\alpha=.93$), \replaced{with}{and subscale reliabilities were} acceptable to excellent \replaced{subscale reliabilities}{(co-presence, attentional allocation, perceived message understanding, and behavioral interdependence: } ($\alpha=.83$--.86). \replaced{Social presence showed no significant effects of \textit{Mode}, \textit{Role}, or their interaction (see~\cref{tab:main_stats_summary}).}{No significant effects were found for \textit{Role} ($F(1,46)=.00$, $p=.33$, $\eta^2_p=.01$), \textit{Mode} ($F(3,138)=2.60$, $p=.06$, $\eta^2_p=.06$), or their interaction ($F(3,138)=1.90$, $p=.14$, $\eta^2_p=.04$).} Ratings were similar across conditions (HU/NU means: \textit{Both} $5.78/5.92$, \textit{H} $5.79/5.93$, \textit{N} $5.72/5.63$, \textit{Off} $5.70/5.59$), suggesting that hand-outline visualization did not substantially affect perceived social presence (\cref{fig:graph_all}-(c)).

\textbf{Preference.}
After the experiment, HUs and NUs ranked the four visualization methods\deleted{ (\textit{Both}, \textit{N}, \textit{H}, \textit{Off}) by preference}. \replaced{Rankings were converted into weighted scores (4--1) and percentages (see~\cref{fig:graph_all}-(d)). For both roles, \textit{Both} was most preferred, followed by \textit{N}, \textit{H}, and \textit{Off}.}{As shown in \cref{fig:graph_all}-(d), rankings were converted into weighted scores (4--1 points from most to least preferred) and percentages. For both roles, \textit{Both} was most preferred (\textit{HU}: 78 points, 34\%; \textit{NU}: 77 points, 35\%), followed by \textit{N} (\textit{HU}: 67, 29\%; \textit{NU}: 61, 28\%), \textit{H} (\textit{HU}: 55, 24\%; \textit{NU}: 52, 24\%), and \textit{Off} (\textit{HU}: 30, 13\%; \textit{NU}: 30, 14\%).}

\section{Discussion}

Overall, our results suggest that the value of tactile-state visualization in asymmetric collaboration lies less in reproducing haptic sensation \deleted{itself }than in restoring shared access to otherwise unavailable object-related tactile information. Cue placement shaped not only task completion time, but also how confidently and actively the non-haptic user (NU) could participate in the joint decision process. \added{We position this work as an initial investigation of roughness visualization as a coarse shared reference for collaborative material judgment, rather than as a fidelity-oriented haptic rendering technique.}

\textbf{Task-relevant hand cues improved efficiency without harming social presence.}
\replaced{For example, Sasaki and Igarashi~\cite{sasaki2025exploring} found that avatar representation improved social presence but not task performance or effort, showing that additional visual information is not inherently beneficial. In contrast, \replaced{showing the cue on both hands reduced completion time
relative to \textit{Off},}{our task-relevant tactile-state visualization reduced completion time} while social presence remained stable across conditions.}{Whereas prior work has shown that avatar visibility primarily influences social presence, with less consistent effects on task performance in collaborative VR~\cite{sasaki2025exploring,wang2025effects,yoon2023effects}, our results suggest a different role for task-relevant tactile-state visualization. \replaced{Compared with the no-cue condition, \textit{Both} reduced task completion time, while social presence remained stable across conditions.}{Compared with the no-cue condition, cue-present conditions improved task completion time, while social presence remained stable across conditions.}} This contrast \deleted{likely}reflects a functional difference between the two cue types: avatar representations mainly support interpersonal awareness, whereas our hand-anchored roughness cue provided directly usable tactile evidence for joint decision-making. In this sense, the cue improved collaboration not by increasing social salience, but by making task-relevant information more accessible. A key implication is that tactile visualization for asymmetric collaboration should prioritize actionable information access over dense or high-fidelity sensory replication.

\textbf{NU-side placement reduced information asymmetry and promoted active participation.}
The results \deleted{further} indicate that placing the cue on the NU's hand reduced the NU's information disadvantage and supported more active involvement in the task. \replaced{When the cue appeared on the NU's hand, the NU had an additional source of task-relevant evidence beyond observing the HU's actions or discussing the cube verbally; the visual proxy itself could support roughness judgment.}{When the cue appeared on the NU's hand, the NU no longer had to rely only on verbal explanation or indirect observation of the HU's actions; instead, the visual proxy itself became evidence for judging roughness.} This interpretation is consistent with the higher confidence and perceived contribution observed in \textit{N} and \textit{Both} than in \textit{Off}. The workload pattern also supports this account: in \textit{Off}, the HU bore a greater interpretive burden than the NU, whereas this imbalance was reduced once tactile-state information became visually accessible. The grasp-time interaction provides additional behavioral evidence. In \textit{H}, grasp time was longer when the HU held the cube, whereas in \textit{N}, grasp time increased when the NU held it, \replaced{suggesting that NU-side placement supported the NU in becoming an \deleted{more} active interpreter of object-related tactile state.}{suggesting that NU-side placement changed the NU from a passive recipient of verbal information into a more active interpreter of object-related tactile state.} Together, these findings suggest that, when only one collaborator has direct haptic access, placing the cue on the deprived partner's hand is an effective way to restore agency and rebalance participation.

\textbf{Shared placement \replaced{showed complementary benefits.}{provided benefits beyond one-sided access recovery.}} \replaced{Both showed a favorable overall pattern: it was the most preferred condition for both roles, reduced completion time relative to Off, and produced lower workload than H. However, we found no clear performance advantage of Both over N; thus, these results do not establish the universal superiority of bilateral placement. One possible explanation is that HU-side visualization provided self-feedback that helped align felt roughness with its visual externalization, while NU-side visualization restored access to tactile-state information. In this sense, shared placement may support both access recovery and mutual grounding. Nevertheless, different contexts may favor different placements: Off may preserve visual clarity, H may support HU self-confirmation, N may support NU access with less visual duplication, and Both may provide a shared reference but introduce clutter or occlusion in visually dense tasks.}{Our results \deleted{also} suggest that showing the cue on both users' hands provided benefits beyond NU-only placement. If the cue functioned only as a one-way aid for the NU, then placing it only on the NU's hand would have been sufficient. Instead, \textit{Both} was the most preferred condition for both roles, produced lower overall workload than \textit{H}, and outperformed \textit{Off} in task completion time. This pattern suggests that shared placement offered more than simple information recovery for the NU. A plausible interpretation is that HU-side placement also acted as self-feedback, helping the HU maintain consistency between felt roughness and its visual externalization during touch. In this sense, mutual placement supported not only information access for the NU, but also alignment and confirmation for the HU. This implies that tactile-state cues in asymmetric collaboration should not be treated solely as one-way transmissions to deprived users, but as shared resources that can support mutual grounding. \added{Although \textit{Both} worked best in our sorting task, it should not be treated as universally optimal. Different contexts may favor different placements: \textit{Off} can preserve visual clarity, \textit{H} can support HU self-confirmation, \textit{N} can support NU access with less visual duplication, and \textit{Both} can provide a shared reference but may introduce clutter or occlusion in visually dense tasks.}}

\textbf{Cue placement should be treated as a collaboration policy rather than a rendering choice.}
Taken together, the performance, workload, confidence, contribution, preference, and grasp-time results suggest that cue placement did not merely alter how information was displayed; it changed how information was distributed and used within the collaboration. In \textit{Off}, tactile information remained available only to the HU. In \textit{N}, access was partially restored to the NU. In \textit{Both}, the cue became a shared reference that supported both access recovery and alignment. The placement decision therefore shaped who could act on tactile evidence, who took a more active role during object handling, and how the interpretive burden was distributed across collaborators. From a design perspective, this means that hand-specific cue placement should be considered a collaboration policy that structures participation and information flow, rather than a purely perceptual rendering parameter.

\textbf{Abstract visual cues can support collaborative tactile judgment.}
Our findings suggest that collaborative tactile support does not necessarily require detailed tactile replication. Even metaphorical visual cues can be sufficient when they provide rapidly interpretable distinctions between coarse tactile categories during  interaction. In our case, the hand-anchored cue did not reproduce texture in a physically faithful way; instead, it offered an abstract visual proxy that allowed collaborators to distinguish roughness categories and use them for joint decision-making. Importantly, our preliminary studies showed that this abstract representation was visually interpretable to non-haptic users~\added{(}\cref{subsec:ps1}\added{)} and meaningfully aligned with roughness perception when combined with haptic input for haptic users~\added{(}\cref{subsec:ps2}\added{)}. This suggests that abstract cross-modal cues can serve as valid collaborative representations of tactile state, and provides a concrete example of how cross-modal cue interpretation may be established for asymmetric collaboration. More broadly, these results suggest that tactile visualizations for asymmetric collaboration may be most effective when they emphasize immediacy, interpretability, and coordination support.

\textbf{Use Cases.} These implications are especially relevant to tasks that require quick, categorical judgments rather than precise material verification. Unlike numeric or bar-style readouts, our hand-anchored cue supports peripheral, contact-situated perception of coarse tactile states. This makes it well suited to early-stage material screening, remote comparison of finishes, online product experience, and haptic playtesting, where collaborators need shared tactile understanding rather than identical tactile fidelity. Accordingly, the approach is likely most useful during early selection and comparison rather than final verification.

\section{Limitations and Future Work}
\label{sec:disc_limitations}

This study has several limitations. First, the experiment involved only dyads with fixed asymmetric roles; scaling to larger groups, role-switching, or heterogeneous devices remains open. Second, we focused on line shape and motion speed as visual proxies for roughness, and it remains unclear how well this approach generalizes to other tactile qualities such as stickiness, softness, or compliance. Third, the study was conducted in a controlled laboratory setting with specific hardware, so future work should examine ecological validity in more realistic collaborative scenarios. \replaced{Fourth, cue placement may be context-dependent; future work should test whether the benefits of \textit{Both}, \textit{H}, or \textit{N} change with object size, visual clutter, or task complexity. Fifth, dyads communicated through natural voice, but we did not systematically record or code verbal exchanges; future work should examine how verbal communication interacts with cue placement and social presence~\cite{merz2024does}. Finally, we did not directly test whether HUs and NUs formed identical subjective roughness percepts for each object; our claim is limited to task-level shared use of the cue during collaborative sorting. Future work should measure pair-level perceptual agreement and explore multimodal extensions such as auditory feedback.}{Finally, although we focused on visual externalization alone, future work could investigate how hand-anchored tactile-state cues interact with other modalities such as auditory feedback, and whether multimodal combinations further improve interpretability, coordination, or shared understanding in asymmetric collaboration.}

\section{Conclusion}
We presented a dynamic hand–outline visualization for conveying roughness in asymmetric VR collaboration. Building on preliminary studies, our results confirmed that \replaced{contour sharpness was the dominant visual cue for roughness, while motion modulated perceived roughness in a shape-dependent manner, supporting three discrete visual–haptic mappings.}{sharp contours and faster motion effectively signal higher roughness, forming the basis for a system that communicates three discrete roughness levels between haptic and non-haptic users.} Our main study further showed that the presence and placement of roughness hand-outline visualizations improved collaboration efficiency, redistributed workload, and enhanced the confidence and contribution of non-haptic users, while maintaining overall accuracy. These findings demonstrate that abstract roughness visualizations can reduce sensory asymmetry and support more balanced participation in VR collaboration. This work introduces a novel approach to tactile visualization and offers design implications for multimodal feedback in collaborative VR. Our findings show that such feedback can reduce sensory asymmetry, foster mutual understanding, and promote equitable participation, key to effective collaboration. Future work should extend these visualizations to finer tactile qualities, multi-user contexts, and applications such as education, training, and design review.

\section*{Disclosure on Gen AI Usage}
In accordance with IEEE guidelines, \replaced{we disclose that ChatGPT was used to generate icon assets, reflect user images in the teaser figure, and assist with English editing/proofreading. All scientific content, experimental design, implementation, data analysis, and conclusions were developed and verified by the authors.}{the authors disclose the use of generative AI tools during the preparation of this manuscript. Chat-GPT were used to generate icon assets for the system framework figure, to reflect user images in the teaser image, and to assist with English editing and proofreading. All scientific content, experimental design, system implementation, data analysis, and conclusions are solely the work of the authors.}

\acknowledgments{%
This work was supported by the Institute of Information \& Communications Technology Planning \& Evaluation (IITP) under the ITRC Program (IITP-2026-RS-2024-00436398) and the Virtual Convergence Support Program to Nurture the Best Talents (IITP-2022(2026)-RS-2022-00156435), both funded by the Korea government (MSIT), and by the Korea Institute for Advancement of Technology (KIAT) grant funded by the Korean government (MOTIE) (RS-2025-02304167, HRD Program for Industrial Innovation).
}

\bibliographystyle{abbrv-doi-hyperref}

\bibliography{template}

\end{document}